\documentclass{iopart}
\usepackage[colorinlistoftodos, prependcaption]{todonotes}
\usepackage[british]{babel}
\usepackage[utf8]{inputenc}
\usepackage[T1]{fontenc}

\expandafter\let\csname equation*\endcsname\relax
\expandafter\let\csname endequation*\endcsname\relax
\usepackage{amssymb,amsfonts}
\usepackage{physics}
\usepackage{graphicx}
\usepackage{cite}
\usepackage{float}
\usepackage{xurl}
\usepackage{wrapfig}
\usepackage{gensymb}
\usepackage{caption}
\usepackage{subcaption}
\usepackage{footnote}
\makesavenoteenv{tabular}
\usepackage{wasysym}
\usepackage{mathtools}
\usepackage{bm} 
\usepackage{dsfont}
\usepackage[pdftex, pdftitle={Article}, pdfauthor={Author}]{hyperref} 
\usepackage{doi}
\usepackage[nobottomtitles]{titlesec} 

\usepackage{subfiles} 

\NewDocumentCommand{\evalat}{sO{\big}mm}{%
	\IfBooleanTF{#1}
	{\mleft. #3 \mright|_{#4}}
	{#3#2|_{#4}}%
}

\begin{document}
\title{Quasiclassical theory of out-of-time-ordered correlators}

\textbf{\author{Thomas R. Michel$^1$, Juan Diego Urbina$^2$ and
		Peter Schlagheck$^1$}
}

\address{$^1$ CESAM Research Unit, University of Liège, 4000 Liège, Belgium}
\address{$^2$ Institut für Theoretische Physik, Universität Regensburg, Regensburg, 93040 Germany}

\date{\today} 

\begin{abstract}
Out-of-time-ordered correlators (OTOCs),  defined via the squared commutator of a time-evolving and a stationary operator, represent observables that provide useful indicators for chaos and the scrambling of information in complex quantum systems.
Here we present a quasiclassical formalism of OTOCs, which is obtained from the semiclassical van Vleck-Gutzwiller propagator through the application of the diagonal approximation. 
For short evolution times, this quasiclassical approach yields the same result as the Wigner-Moyal formalism, i.e., OTOCs are classically described via the square of the Poisson bracket between the two involved observables, thus giving rise to an exponential growth in a chaotic regime. 
For long times, for which the semiclassical framework is, in principle, still valid, the diagonal approximation yields an asymptotic saturation value for the quasiclassical OTOC under the assumption of fully developed classical chaos.
However, numerical simulations, carried out within chaotic few-site Bose-Hubbard systems in the absence and presence of periodic driving, demonstrate that this saturation value strongly underestimates the actual threshold value of the quantum OTOC, which is normally attained after the Ehrenfest time.
This indicates that nondiagonal and hence genuinely quantum contributions, thus exceeding the framework of the quasiclassical description, are primarily responsible for describing OTOCs beyond the short-time regime.
\end{abstract}

\maketitle

\section{Introduction}
\indent Quantum chaos, referring to the characteristic quantum signatures that are generally exhibited by systems having a chaotic classical counterpart \cite{Haake}, has gained a lot of traction in recent years \cite{Richter_Urbina_Tomsovic_2022}.
Originally rooted in nuclear and atomic physics \cite{Bohigas_Weidenmueller_1988}, the subject is now related to many other fields such as mesoscopic physics \cite{Beenakker97RMP}, optics and microwave physics \cite{Stockmann_1999}, many-body physics \cite{Sachdev_Ye_1993, Nandkishore_Huse_2015}, as well as black-hole physics \cite{Sekino_Susskind_2008, Lashkari_Stanford_Hastings_Osborne_Hayden_2013, Shenker_Stanford_2014, Hayden_Preskill_2007} and quantum information \cite{Hayden_Preskill_2007}. 
Classical chaos being characterized by the exponential spreading of trajectories that are started nearby in phase space, the corresponding quantum counterpart is the spreading of information, also called information scrambling, leading eventually to thermalisation. 

Useful indicators of these phenomena are Out-Of-Time-Ordered Correlators, or OTOCs \cite{LarOvc69,Swingle_2018, Garcia_Bu_Jaffe_2023}, which can be applied to a wide range of systems, including simple models as well as complex many-body systems.
Similarly as for the Loschmidt echo \cite{Jalabert_Pastawski_2001}, the concept of OTOCs allows, in a semiclassical context, for the appearance of the classical Lyapunov exponent in the framework of a quantum observable.
More specifically, for systems with a chaotic classical counterpart OTOCs can be shown to grow exponentially for short times, with the exponential growth rate being determined by the maximum Lyapunov exponent of the system \cite{Maldacena_Shenker_Stanford_2016, Rozenbaum_Ganeshan_Galitski_2017, Rammensee_Urbina_Richter_2018}.
This exponential growth is not unlimited but turns into a saturation of the OTOC observable beyond the Ehrenfest time $t_E$ of the chaotic system.
It can be reproduced, but without saturation, by means of a quasiclassical approach based on the Wigner-Moyal formalism, which is strictly perturbative in $\hbar$ and not suited for being applied for time scales larger than $t_E$.
Proper semiclassical reasoning, applied to generic systems with chaotic behaviour, allows one to quantitatively explain the occurrence of saturation in terms of trajectory pairs that exhibit low-angle crossing \cite{Rammensee_Urbina_Richter_2018, Rammensee_2019}.
In a very similar manner as for loop corrections to the spectral form factor \cite{SieberRichter01}, those trajectory pairs give rise to contributions to the OTOC that are intrinsically nonclassical, in the sense that they do not arise in the framework of a diagonal approximation \cite{Berry85} and thus cannot be incorporated into a quasiclassical numerical simulation scheme based on the Truncated Wigner approximation \cite{Steel98PRA,Sun98JCP}.

The present paper aims at bridging the gap between the Wigner-Moyal approach limited to short times and the generic semiclassical approach of \cite{Rammensee_Urbina_Richter_2018} which is valid for short and long times but less useful for numerical simulations.
To this end, we develop a full-fledged semiclassical theory of OTOCs based on the van Vleck-Gutzwiller propagator \cite{Gutzwiller67,Gutzwiller90}, which is then used to derive a quasiclassical description via a diagonal approximation.
The approach presented here can be seen as an extension (and, in some key aspects, a correction) of two previous attempts towards this aim \cite{Jalabert_2018,Kurchan_2018}.
In \cite{Jalabert_2018} both short- and long-time behaviour of the OTOC are calculated in the framework of the van Vleck-Gutzwiller propagator, using extra assumptions that extend beyond the bare semiclassical propagator. 
For the short time regime, the propagation is approximated by free evolution (a step justified only for billiard systems) while ergodic arguments are invoked to explain the saturation at long times. 
Similarly, the calculation presented in \cite{Kurchan_2018} bypasses these restriction for the short time dynamics by addressing only the dynamics of a particle on a Riemannian manifold of constant negative curvature, which is again a particular type of chaotic system. 

Avoiding any additional assumption, our calculation yields a quasiclassical description of OTOCs that is valid for general hamiltonian systems both in the short-time and in the long-time regime.
While in the short-time regime the exponential increase of the OTOC is recovered, exactly as predicted by the Wigner-Moyal approach, our quasiclassical framework allows us also to obtain a prediction for an asymptotic long-time value of this observable.
However, comparisons with numerical simulations, which are carried out in the framework of periodically driven Bose-Hubbard dimers and undriven Bose-Hubbard trimers, reveal that this particular long-time value strongly underestimates the true threshold value of the OTOC.
We conclude from this finding that the genuinely nonclassical contributions stemming from trajectory pairs with low-angle crossing play a dominant role for times beyond the Ehrenfest time, as was also conjectured in \cite{Rammensee_Urbina_Richter_2018}.

The paper is structured as follows. 
Section \ref{sec_OTOTCinChaoticSystems} presents the state of the art of OTOCs in chaotic systems. 
We give here a brief account on the Wigner-Moyal approach and explain the general ideas behind the reasoning developed in \cite{Rammensee_Urbina_Richter_2018}.
In Section \ref{sec_DiagonalApprox} we present the general quasiclassical expression for the OTOC as it is derived from the semiclassical van Vleck-Gutzwiller propagator through the application of the diagonal approximation. 
Details of these calculations are provided in the appendices.
In Sections \ref{sec_SemiclassicalLimit} and \ref{sec_nohbarExpansion} we respectively discuss the short-time and the long-time limit of the quasiclassical OTOC. 
Comparisons with numerical simulations carried out within Bose-Hubbard systems are presented and discussed in Section \ref{sec_numerics}.

\section{Out-of-time-ordered correlator in chaotic systems} \label{sec_OTOTCinChaoticSystems}
\indent Out-of-time-ordered correlators can be defined as the expectation value of the square modulus of a commutator where one of the operator is time-evolved:
\begin{eqnarray} \label{OTOC_def}
    C(t) = \bra{\psi} \abs{\left[\hat{A}(t), \hat{B}(0) \right]}^2 \ket{\psi}.
\end{eqnarray}
Here $\hat{A}$, $\hat{B}$ are local operators, $\hat{A}(t)$ is the time-evolved operator $\hat{A}$ in the Heisenberg representation, given by
\begin{eqnarray} \label{A_t_operator}
    \hat{A}(t) =  \hat{U}^\dagger(t) \hat{A} \hat{U}(t),
\end{eqnarray}
with $\hat{U}(t)$ the evolution operator, and $\ket{\psi}$ is the initial state of the system \cite{Rammensee_Urbina_Richter_2018}. The operators are assumed hermitian to simplify the derivations.\\
\indent The formalism that we developed is applicable to generic one- or multi-dimensional quantum systems that possess a classical limit. A good candidate is the previously mentioned Bose-Hubbard system, described in more details in Section \ref{sec_numerics}. It consists of sites where bosons can hop from one site to an adjacent one and interact in a 2-body fashion on each site. In this system, examples of hermitian operators are the quadrature momentum and position operators, respectively $\hat{q}_i = (\hat{b}_i+\hat{b}^\dagger)/\sqrt{2}$ and $\hat{p}_i = (\hat{b}_i-\hat{b}^\dagger)/\sqrt{2i}$, with $\hat{b}_l$ and $\hat{b}^\dagger_l$ the annihilation and creation operators on site $l$, as well as population operators $\hat{n}_l = \hat{b}^\dagger_l \hat{b}_l$.\\
\indent Since our goal is to look at the classical limit, a naturally-arising tool is the Wigner-Weyl formalism. It is a phase-space formulation of quantum mechanics which allows us to work with smooth phase-space functions instead of operators. Those, as well as states, are replaced by objects called respectively Weyl symbols and Wigner functions. For any operator $\hat{A}$, one defines its Weyl symbol as
\begin{eqnarray}
    A(\bm{q},\bm{p}) \equiv (\hat{A})_W(\bm{q},\bm{p})= \int \dd\xi\ \bra{\bm{q}+\frac{\bm{\xi}}{2}} \hat{A} \ket{\bm{q}-\frac{\bm{\xi}}{2}} \rme^{-\rmi \bm{p \cdot \xi}/\hbar}
\end{eqnarray}
 with $\bm{q}$, $\bm{p}$, $\bm{\xi}$ phase-space vectors. Additionally, the symbol of a product of operators is given by the Moyal product
\begin{eqnarray}
    (\hat{A}\hat{B})_W(\bm{q},\bm{p}) &= A(\bm{q},\bm{p}) \exp(\rmi \hbar \sum_l^L \left(\overleftarrow{\partial_{q_l}} \overrightarrow{\partial_{p_l}} - \overleftarrow{\partial_{p_l}} \overrightarrow{\partial_{q_l}}\right)/2 ) B(\bm{q},\bm{p}) 
\end{eqnarray}
where the arrows indicate the side on which the derivative operators are applied \cite{Polkovnikov_2010}. It follows that the symbol of a commutator is 
\begin{eqnarray}
    \left[\hat{A},\hat{B}\right]_W(\bm{q},\bm{p}) & = 2 A(\bm{q},\bm{p}) \sin(\frac{\hbar}{2} \sum_l^L \overleftarrow{\partial_{q_l}} \overrightarrow{\partial_{p_l}} - \overleftarrow{\partial_{p_l}} \overrightarrow{\partial_{q_l}}) B(\bm{q},\bm{p}).
\end{eqnarray}
where $\left[\cdot,\cdot \right]_W$ is called the Moyal bracket. The Wigner function is, up to a global prefactor, given by the Weyl symbol associated with the density operator of the system. In the case of a pure state, it is written as
\begin{eqnarray}
    W(\bm{q},\bm{p}) \equiv \frac{1}{\left(2\pi \hbar\right)^L} \int \dd\bm{\xi}\ \psi^*\left(\bm{q}+\frac{\bm{\xi}}{2}\right) \psi\left(\bm{q}-\frac{\bm{\xi}}{2}\right) \rme^{\rmi \bm{p}\cdot\bm{\xi}/\hbar}.
\end{eqnarray}
This Wigner-Moyal formalism can be used to obtain the short-time approximation of the OTOC, which consists in a direct $\hbar \to 0$ limit. Mathematically, this amounts to replacing the commutator of the quantum operators of the OTOC definition by a Poisson bracket of the corresponding Weyl symbols, multiplied by $\rmi \hbar$. Using for instance $\hat{A} = \hat{q}_i$, $\hat{B} = \hat{p}_j$, this becomes
\begin{eqnarray}
    \bra{\psi} \left[\hat{q}_i(t), \hat{p}_j \right]^2 \ket{\psi} & =  \int \dd \bm{q} \dd\bm{p}\ \left(\left[\hat{q}_i(t), \hat{p}_j \right]^2\right)_W(\bm{q},\bm{p})\ W(\bm{q},\bm{p}) \nonumber           \\
                                                                  & \to \hbar^2 \int\dd \bm{q} \dd\bm{p}\ \{ q_i(t), p_j\}^2\ W(\bm{q},\bm{p}) \nonumber                                                      \\
                                                                  & = \hbar^2 \int\dd \bm{q} \dd\bm{p}\ \left( \frac{\partial q_i}{\partial q_j}(t)  \right)^2 W(\bm{q},\bm{p}) \propto \hbar^2 e^{2\lambda t}
\end{eqnarray}
where $\lambda$ is the classical Lyapunov exponent of the system \cite{Richter_Urbina_Tomsovic_2022}. The exponential growth holds provided that the system is chaotic. This approximation is valid until the Ehrenfest time, where the OTOC saturates due to interferences. Wigner-Moyal does not account for this effect but results in a never-ending exponential growth.\\
\indent  In order to obtain an expression that is not necessarily time-restricted, we use the semiclassical approximation known as the \emph{van Vleck-Gutzwiller propagator}, which consists in performing a stationary-phase approximation of the Feynman path integral \cite{Richter_Urbina_Tomsovic_2022}. It expresses the propagator appearing in the path integral as a coherent sum over all classical paths linking the departure $\bm{q}^{\rm{i}}$ and arrival points $\bm{q}^{\rm{f}}$ in a given time $t$, according to
\begin{eqnarray} \label{vVG}
    K(\bm{q}^{\rm{f}}, \bm{q}^{\rm{i}}, t) & \equiv \bra{\bm{q}^{\rm{f}}} \hat{U}(t) \ket{\bm{q}^{\rm{i}}} \simeq \sum_{\gamma:\bm{q}^{\rm{i}}\to \bm{q}^{\rm{f}}} A_\gamma (\bm{q}^{\rm{f}}, \bm{q}^{\rm{i}}, t)\ \rme^{\rmi R_\gamma(\bm{q}^{\rm{f}}, \bm{q}^{\rm{i}}, t)/\hbar}
\end{eqnarray}
with 
\begin{eqnarray} A_\gamma (\bm{q}^{\rm{f}}, \bm{q}^{\rm{i}}, t) &= \frac{1}{(2\pi \hbar)^{L/2}} \abs{\det \frac{\partial^2 R_\gamma}{\partial \bm{q}^{\rm{f}} \partial \bm{q}^{\rm{i}}} (\bm{q}^{\rm{f}},\bm{q}^{\rm{i}},t)}^{1/2}  \rme^{- \rmi \mu_\gamma \pi/4} \nonumber\\
R_\gamma(\bm{q}^{\rm{f}}, \bm{q}^{\rm{i}}, t) &= \int_{0}^{t} \dd t^\prime  \left( \bm{p}_\gamma(t^\prime) \cdot \bm{\dot{q}}_\gamma (t) - H_{cl} (\bm{q}_\gamma (t), \bm{p}_\gamma (t))\right)
\end{eqnarray}
respectively the amplitude of the propagator and the action of the trajectory. $\mu_\gamma$ is the Maslov index related to the trajectory $\gamma$, and $H_{cl}(\bm{q}, \bm{p}) \equiv ( \hat{H})_W (\bm{q},\bm{p})$. This approximation preserves the coherent aspect of quantum mechanics \cite{Rammensee_Urbina_Richter_2018,Richter_Urbina_Tomsovic_2022,Engl_Urbina_Richter_2016}. In the following, as the amplitude appears inside a modulus, Maslov's indices will not be present anymore.
\\
\indent Our work follows what has been done in \cite{Rammensee_Urbina_Richter_2018}, although in a slightly different fashion. In Rammensee \textit{et al.}'s paper, the path-integral is expressed as a coherent sum over mean-field solutions of the time evolution. Then, the dominant many-body interference terms are computed., which results in contributions that can be represented by 4 diagrams, as in shown in \fref{fig_pairing}. Quantum interference is thus considered to some degree in so-called \textit{encounter regions}, which are regions where the four trajectories are in the vicinity of each other such as in the leftmost part of \fref{family1}.\\
\indent In our work, we start from the same path integral but restrict ourselves to strict quasiclassical contributions which do not involve any interference term. Both formalisms predict a finite and non-trivial long-time value. \cite{Rammensee_Urbina_Richter_2018} identifies two contributions, \fref{family2} and \fref{family4}, whereas our formalism does not enable us to capture the latter. This allows to identify quantitatively the importance of interference, as our long-time value is purely quasiclassical.

\newcommand{\trajfigwidth}{0.55}
\begin{figure}[t]
    \centering
    \begin{subfigure}[t]{.4\textwidth}
        \centering
        \includegraphics[scale=\trajfigwidth, trim={0cm 0cm 0cm 0cm}, clip]{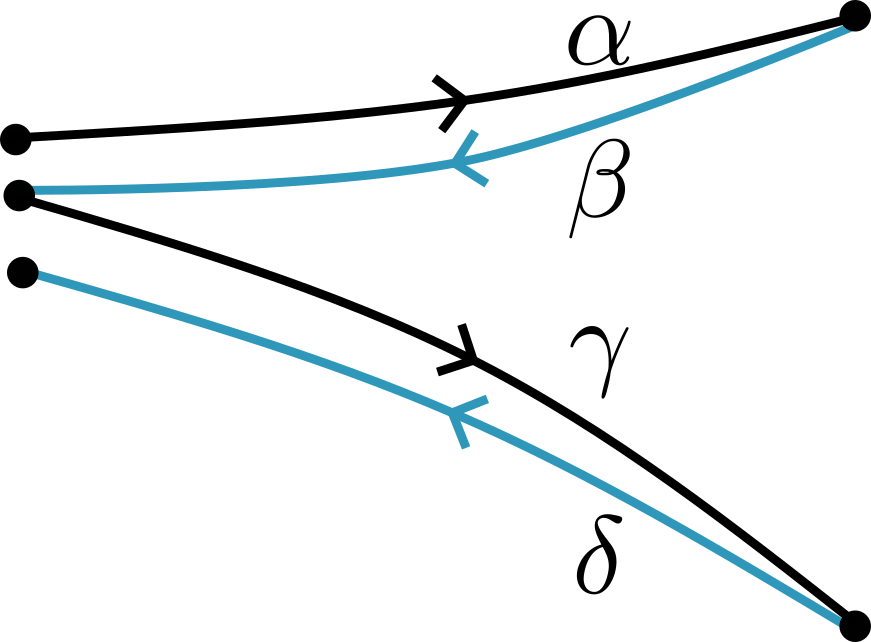}
        \caption{V pairing}
        \label{family1}
    \end{subfigure}
    \begin{subfigure}[t]{.4\textwidth}
        \centering
        \includegraphics[scale=\trajfigwidth, trim={0cm 0cm 0cm 0cm}, clip]{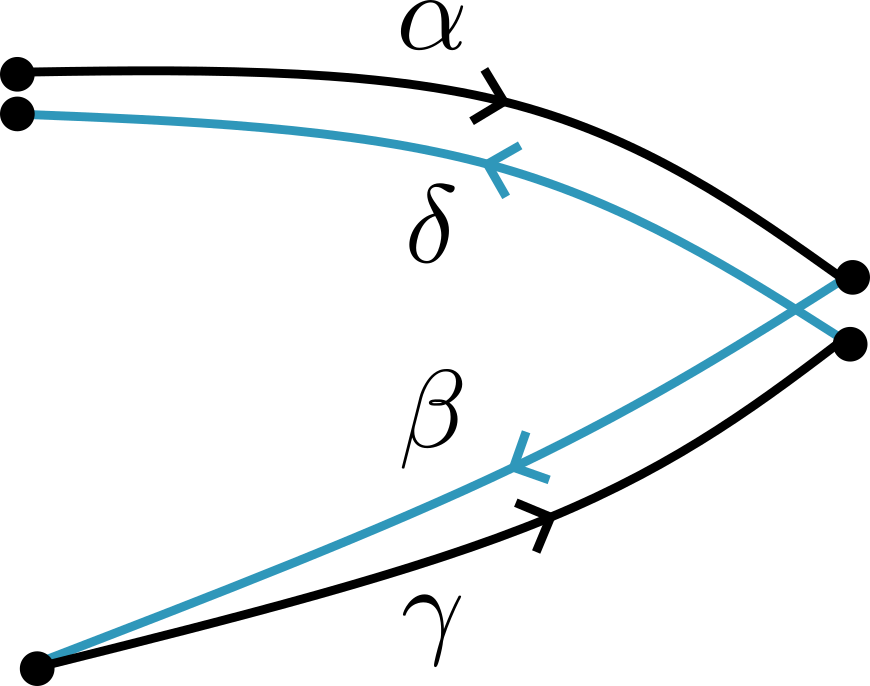}
        \caption{$\Lambda$ pairing}
        \label{family2}
    \end{subfigure}
    \newline
    \begin{subfigure}[t]{.4\textwidth}
        \centering
        \includegraphics[scale=\trajfigwidth, trim={0cm 0cm 0cm 0cm}, clip]{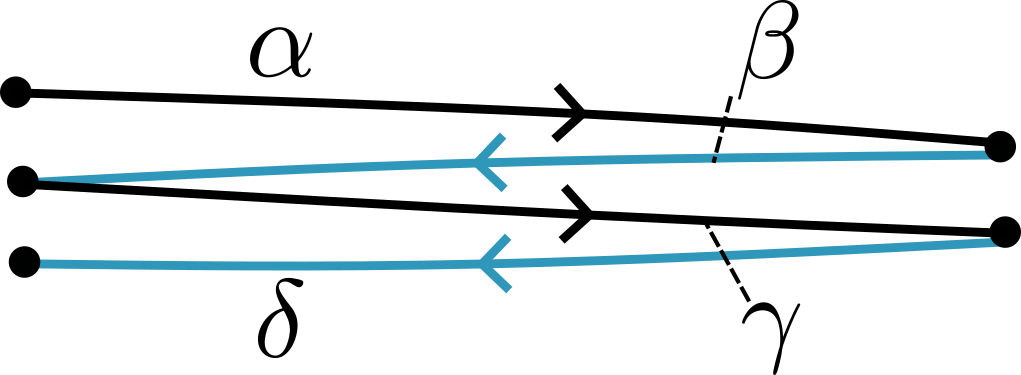}
        \caption{I pairing}
        \label{family3}
    \end{subfigure}
    \begin{subfigure}[t]{.4\textwidth}
        \centering
        \includegraphics[scale=\trajfigwidth, trim={0cm 0cm 0cm 0cm}, clip]{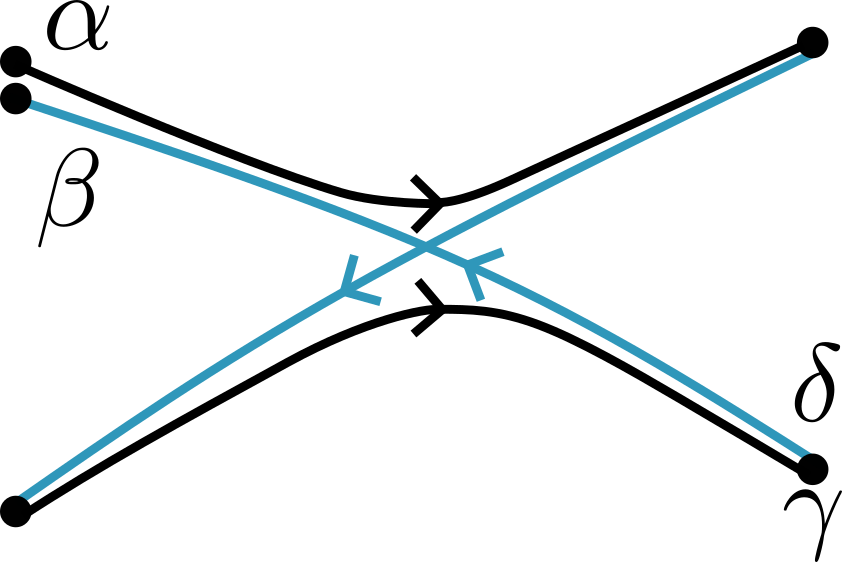}
        \caption{X pairing}
        \label{family4}
    \end{subfigure}
    \caption{The four contributions to the OTOC, which (a) $V$ pairing, (b) $\Lambda$ pairing, (c) $I$ pairing and (d) $X$ pairing, labelled according to their shape. Time axis is assumed horizontal, with the future on the right. The black trajectories ($\alpha,\ \gamma$) are forward-propagating, whereas the blue trajectories ($\beta,\ \delta$) are backward-propagating.}
    \label{fig_pairing}
\end{figure}

\section{Diagonal approximation of out-of-time-ordered correlators} \label{sec_DiagonalApprox}
All the terms from the OTOC definition \eqref{OTOC_def} can be explicitely written:
\begin{flalign}
    C(t) 
	&=  \bra{\psi} \hat{B}\hat{A}(t)\hat{A}(t)\hat{B} - \hat{B}\hat{A}(t)\hat{B}\hat{A}(t)- \hat{A}(t)\hat{B}\hat{A}(t)\hat{B} + \hat{A}(t)\hat{B}\hat{B}\hat{A}(t)\ket{\psi}.&
\end{flalign}
Our goal is to evaluate the classical limit of this OTOC, and our first step is the semiclassical propagator. We make the quantum propagator appear via $\hat{A}(t)$ from \eqref{A_t_operator} and by using \eqref{vVG}. More details can be found in \ref{sec_app_GenericTerm}. Four propagators appear, meaning that the phase contains combinations of actions related to four trajectories. One can make the reasonable assumption that uncorrelated paths, with uncorrelated actions, translate into highly oscillatory terms that average out. In other words, the only non-vanishing contributions to the integral are paths that have correlated actions. A particular case of such an approximation is the \emph{diagonal approximation}. Inside the phase, the action of a forward-propagating (respectively a backward-propagating) trajectory appears with a plus (respectively a minus) sign. The diagonal approximation consists in identifying pairs of forward- and backward-propagating trajectories that result in a vanishing action difference at the $0^{\text{th}}$ order for a given pair. Contrarily to the semiclassical approximation, doing so gets rid of all considerations of interference and renders the theory classical. Nonetheless, as no hypothesis on $\hbar$ has been made thus far, we will refer to this framework as \textit{quasiclassical}. It also means that this approximation can in principle be valid beyond the Ehrenfest time. The three resulting pairings of the diagonal approximation are shown in figure \ref{family1}, \ref{family2} and \ref{family3} and are named according to their respective shape. We will be calling them the $V$, $\Lambda$, $I$ pairings. In order to go further and account for quantum effects, the first correction is the $X$ pairing, represented by \ref{family4} \cite{Rammensee_Urbina_Richter_2018}.\\
\indent The quasiclassical OTOC is written as a sum of three contributions:
\begin{eqnarray} \label{Ccl_sumcontrib}
	C_{qcl(t)} = C^V(t) + C^\Lambda(t) -C^I(t).
\end{eqnarray}
As the $I$ pairing is already contained within the $V$ and $\Lambda$ pairing, we must remove the overcounting, hence the minus. Calculations are done in \ref{sec_app_GenericTerm}-\ref{sec_app_OTOC_allpairings}.

\indent Let us first introduce a more compact way of writing expressions: the symplectic formalism. It exploits the symmetry between the position and momentum arguments and consists in writing the math in terms of phase-space vectors, which we will call symplectic vectors, using the symplectic (or wedge) product $\wedge$ defined as follows: if $X = (\bm{q}_1, \bm{p}_1)$ and $Y = (\bm{q}_2, \bm{p}_2)$, then
\begin{eqnarray}
X \wedge Y = X J Y =\bm{q}_1 \cdot \bm{p}_2 - \bm{q}_2 \cdot \bm{p}_1,    
\end{eqnarray} with $J$ the symplectic matrix \cite{Brodier_Almeida_2004}.\\
\indent The $V$ pairing considers that the families of trajectories $\alpha$ and $\beta$ (respectively $\gamma$ and $\delta$) are one and the same. This is written $\alpha \sim \beta$ (respectively $\gamma \sim \delta$).
Using this symplectic notation, we obtain the quasiclassical expression for the $V$ pairing of the OTOC:
\begin{flalign} \label{OTOC_pair1}
    C^V(t) =\ &\frac{1}{(2\pi)^{4L}} \int \dd X_{1,2,3} \dd\Delta_{1,2}\ \rme^{\rmi\Delta_1 \wedge (X_2 - X_3)} \rme^{\rmi \Delta_2 \wedge (X_1 - 2X_2 + X_3) } &\nonumber\\
    &\times A_t\left(X_1 + \frac{\hbar \Delta_2}{4}\right) A_t\left(X_1 - \frac{\hbar \Delta_2}{4}\right) \biggl[\biggr. B\left(X_2 +  \frac{\hbar\Delta_1}{4}\right)\ B\left(X_2 -  \frac{\hbar\Delta_1}{4}\right) &\nonumber\\
    &  - B\left(2 X_1 -2 X_2 + X_3\right) \left(B\left(X_2 +  \frac{\hbar\Delta_1}{4}\right)+ B\left(X_2 -  \frac{\hbar\Delta_1}{4}\right)\right) &\nonumber\\
	&  + (\hat{B}^2)_W\left(2 X_1 -2 X_2 + X_3\right) \biggl.\biggr] W(X_3)&
\end{flalign}
where $A_t(X) \equiv A(X_t(X))$ is the symbol of $\hat{A}$ evaluated at the time-evolved coordinate $X_t(X)$, often (abusively) denominated the time-evolved symbol. The OTOC has thus been rewritten as an initial value problem in the form of a phase space integral weighed by the Wigner function.


\indent Concerning the $\Lambda$ term, the main difference with the $V$ term lies in the fact the different pairing of trajectories leads to different center-of-mass and relative coordinates. Nonetheless, the steps remain the same and we obtain
\begin{flalign} \label{OTOC_pair2}
	C^\Lambda(t)(t) =\ &\frac{1}{(2\pi)^{4L}} \int \dd X \dd S_{2,3} \dd \Delta_{1,2}\ \rme^{\rmi \Delta_1\wedge(X_{-t}(2S_2-S_3)-X)} \rme^{\rmi \Delta_2\wedge (S_3 - S_2)} &\nonumber\\
    &\times A\left(S_2+ \frac{\hbar \Delta_2}{4}\right) A\left(S_2- \frac{\hbar \Delta_2}{4}\right) \biggl[\biggr. B\left(X+\frac{\hbar \Delta_1}{4}\right) B\left(X-\frac{\hbar \Delta_1}{4}\right) &\nonumber\\
	&-B\left(X_{-t}(S_3)\right)\left(B\left(X+\frac{\hbar \Delta_1}{4}\right) + B\left(X-\frac{\hbar \Delta_1}{4}\right)\right)  &\nonumber\\
	&+\left(\hat{B}^2\right)_W\left(X_{-t}(S_3)\right)\biggl.\biggr]  W\left(2X-X_{-t}(2S_2-S_3)\right).&
\end{flalign}
where $X_{-t}(S)$ means the point time-evolved backward\footnote{The evolution starts at time $t$ and ends at time $0$.} from final conditions $S$.


In the $I$ term term, the four trajectories are assumed to remain within the vicinity of each other at all time. One would then be tempted to change to some global center-of-mass and relative coordinates. It is however more advisable to keep the same variables as for the $\Lambda$ term. 
Since there is only one family of trajectories, there is not a double sum $\sum_{\alpha,\gamma} \abs{A_\alpha}^2 \abs{A_\gamma}^2$ as prefactor, but rather $\sum_\alpha \abs{A_\alpha}^2 \abs{A_\alpha}^2$, which requires extra steps to be dealt with.

We obtain a very similar expression to \eqref{OTOC_pair2}:
    \begin{flalign} \label{OTOC_pair3}
    C^I(t) =\ &\frac{1}{(2\pi )^{4L} } \int \dd X \dd S_{2,3} \dd \Delta_{12} \ \delta_{\alpha_2,\alpha_3} \rme^{\rmi \Delta_1 \wedge (X_{-t}(2S_2-S_3)-X)} \rme^{\rmi \Delta_2 \wedge (S_3-S_2)}   &\nonumber\\
	& \times A\left(S_2 + \frac{\hbar \Delta_2}{4}\right) A\left(S_2- \frac{\hbar \Delta_2}{4}\right)  \biggl[\biggr. B\left(X+\frac{\hbar \Delta_1}{4}\right) B\left(X-\frac{\hbar \Delta_1}{4}\right) &\nonumber\\ 
	&  -B\left(X_{-t}(S_3)\right)\left(B\left(X+\frac{\hbar \Delta_1}{4}\right) + B\left(X-\frac{\hbar \Delta_1}{4}\right)\right) +\left(\hat{B}^2\right)_W\left(X_{-t}(S_3)\right) \biggl.\biggr] &\nonumber\\
    & \times W\left(2X-X_{-t}(2S_2-S_3)\right) &
\end{flalign}
where the only difference is the Kronecker delta $\delta_{\alpha_2,\alpha_3}$ which accounts for the fact that the trajectories that end at $S_2$ and $S_3$ must belong to the same family, and thus remain in the vicinity of each other during the whole evolution. This additional Kronecker will be $1$ for fixed time as $\hbar \to 0$. Nonetheless, as time gets larger, the initial conditions must be closer to each other in order to contribute, meaning that the contribution to \eqref{OTOC_pair3} will get smaller, down the point where it vanishes. Physically, this should be understood as the fact that all trajectories remaining in the vicinity of each other during the whole time evolution becomes increasingly unlikely.

\section{Classical limit for short time} \label{sec_SemiclassicalLimit}
Let us now discuss the classical limit $\hbar \to 0$ of the OTOC expression \eqref{Ccl_sumcontrib}. We first focus on the $V$ pairing, where more details can be found in \ref{sec_app_OTOC_Vpairing}. 
%
We start from \eqref{OTOC_pair1}. The $0^\text{th}$ order consists in neglecting any $\hbar$ dependence. By doing so, we end up with $\Delta_1$ and $\Delta_2$ appearing only inside the complex exponential. Performing the corresponding integrals yields two Dirac delta's, resulting in $X_1=X_2=X_3$, which make the parenthesis with the $B$ symbols vanish: $C^V(t) = 0 + \mathcal{O}(\hbar^1)$.
%
We then easily show that the $1^\text{st}$ order is zero due to the symmetry $\hbar \leftrightarrow -\hbar$ and to the fact that the 1$^\text{st}$ order of the symbol of a square operator vanishes\footnote{See Appendix \ref{app_sqopp}}: $C^V(t) = 0 + \mathcal{O}(\hbar^2)$.\\
%
\indent The first non-vanishing contribution to the OTOC, which we call the \emph{classical OTOC} of the $V$ pairing $C^V_{cl}$, is the $2^\text{nd}$ order in $\hbar$. We obtain
\begin{flalign} \label{math_OTOC2ndorder}
	C^V_{\rm{cl}}(t) &= \hbar^2 \int \dd X \sum_{k,l=1}^{2L} \sigma_k \sigma_l \frac{\partial A_t}{\partial X_{\bar{k}}}(X)\ \frac{\partial A_t}{\partial X_{\bar{l}}}(X)\ \frac{\partial B}{\partial X_k}(X)\ \frac{\partial B}{\partial X_l}(X)\ W(X) & \nonumber\\
	&= \hbar^2 \int \dd X \left(\sum_{k=1}^{2L} \sigma_k \frac{\partial A_t}{\partial X_{\bar{k}}}(X)\  \frac{\partial B}{\partial X_k}(X) \right)^2 W(X) &
\end{flalign}
with $\sigma_i = 1$ and $\bar{i}=i+L$ if $i \in \{1,...,L \}$, and $\sigma_i=-1$ and $\bar{i}=i-L$ if $i \in \{ L+1,..., 2L\}$.  This $\sigma_i$ variable was introduced purely to be able to write everything under the same summation sign. It's important to note that simply doing an $\hbar$ expansion yields an expression that is not equivalent to \eqref{math_OTOC2ndorder} at first sight. One has to use different integrations by parts to tweak the expression as needed.\\
\indent When switching from the symplectic notation back to the regular one, \textit{i.e.} with indices running from 1 to $L$, we obtain a Poisson bracket:
\begin{flalign}
	C^V_{\rm{cl}}(t) &= \hbar^2 \int \dd \bm{q} \dd\bm{p} \left(\sum_{k=1}^{L}  \frac{\partial A_t}{\partial p_k}(\bm{q},\bm{p})  \frac{\partial B}{\partial q_k}(\bm{q},\bm{p}) - \frac{\partial A_t}{\partial q_k}(\bm{q},\bm{p})  \frac{\partial B}{\partial p_k}(\bm{q},\bm{p}) \right)^2 W(\bm{q},\bm{p})  & \nonumber\\
	& = \hbar^2 \int \dd \bm{q} \dd\bm{p} \left\{ A_t(\bm{q},\bm{p}), B(\bm{q},\bm{p})\right\}^2 W(\bm{q},\bm{p}) &
\end{flalign}
which is exactly what is obtained using the Wigner-Moyal formalism. The main but crucial difference is the way we obtained this expression: we used the semiclassical propagator followed by the diagonal approximation, and only then did we make an $\hbar$ expansion to obtain the squared Poisson bracket. Doing so offers more control in the path towards the classical limit and can in principle, if one either restrains from doing expansions or only do them outside evolution operators, yield a quasiclassical formulation of the OTOC valid for all time \cite{Sepulveda_Tomsovic_Heller_1992}.

The same process can be done for the $\Lambda$ and $I$ pairings and the same result is obtained in both cases for short times:
\begin{flalign}
    C^\Lambda_{cl}(t) &= C^I_{cl}(t) &\nonumber\\
    &=\hbar^2 \int \dd X\ \left\{ A(S(X)), B(X)\right\}^2 W(X) \nonumber\\
    &\quad - \frac{\hbar^2}{2} \int \dd S \sum_{i,j=1}^{2L} \sigma_i \sigma_j  A(S) \partial^2_{ij} A(S) \frac{\partial B}{\partial S_{\bar{i}}} (X_{-t}(S)) \frac{\partial B}{\partial S_{\bar{j}}}(X_{-t}(S))  W(X_{-t}(S)) &\nonumber\\
    &\quad + \frac{\hbar^2}{2} \int \dd X \sum_{i,j=1}^{2L} \sigma_i \sigma_j  A(S) \frac{\partial^2 A}{\partial X_{\bar{i}} \partial X_{\bar{j}}} (X) \partial_i B(X) \partial_j B(X) W(X)&
\end{flalign}
The details and are presented in \ref{sec_app_OTOC_Lambdapairing} and \ref{sec_app_OTOC_Ipairing}. The first non-vanishing contribution is again of order $\hbar^2$ and we can recover the Poisson bracket plus some additional terms. As the same result is obtained for short times, the two pairings cancel each other and only the $V$ pairing contributes to the short-time classical OTOC.

\section{No $\hbar$ expansion: a long-time limit} \label{sec_nohbarExpansion}
In this section, we derive a finite long-time limit of the quasiclassical OTOC. In chaotic systems, any small perturbation grows exponentially with time and becomes large. Here, we refrained from doing $\hbar$ expansion involved inside an evolution operator as it is incompatible with considering the system at time larger than the Ehrenfest time. We also considered only the $V$ pairing for two reasons. The first is that the $I$ pairing is expected not to contribute at long times, as the fact that all four trajectories remaining in the vicinity of each other at all time in a chaotic system is not possible by definition. The second is that we expect the $\Lambda$ pairing to yield a similar contribution to $C^V$ for long times.
We make the hypothesis that our system is ergodic and mixing. In the case of a fully-chaotic phase space, this is not a restrictive statement. We will see later how this can be generalised to mixed phase space.  One consequence is that the expression that replaces $A_t$ will depend only on the constants of motions. Assuming $K$ constants of motion:
\begin{eqnarray}
	A_t(X) &\to \bar{A}(\vec{c}(X)) = \int \dd X^\prime \ A(X^\prime) \prod_{k=1}^K \delta(c_k(X^\prime)-c_k(X))
\end{eqnarray}
where $\bar{A}$ means the phase-space average of $A_t$ and $c_k(X)$ is the $k^\text{th}$ constant of motion evaluated at $X$.\\
\indent To explain the average, let us consider an integral of the form $I(t)= \int \dd x\ f_t(x) g(x)$, where $f_t$ involves some time evolution and varies much more quickly than $g$ with $x$. One can divide the $x$ space in cells and compute the integral cell by cell. This would translate to $I\simeq \sum_{\text{cell}_i} g(\text{cell}_i)\int_{\text{cell}_i} f_t(x) \dd x = \sum_{\text{cell}_i} g(\text{cell}_i) \bar{f}(\text{cell}_i)$, where $g$ could be taken out of the integral on a given cell as it varies slowly, and where $\bar{f}$ is the average of $f_t$ on the whole space because of the mixing. One can then take the limit of infinitely small cells, which gives back the integration over the whole space: $I = \int \dd x\ \bar{f}(x) g(x)$.
Moreover, because of ergodicity, it is also equal to the time average.\\
\indent Using this in \eqref{OTOC_pair1} translates to
\begin{eqnarray}
	\fl A_t\left(X+\frac{\hbar \Delta_2}{4}\right) &\to& \bar{A}\left(\vec{c}\left(X+\frac{\hbar \Delta_2}{4}\right)\right) \\
	&=&\bar{A}\left(\vec{c}\left(X\right)\right) + \frac{\hbar }{4}\sum_{m=1}^K \sum_{i=1}^{2L} \Delta_{2,i} \frac{\partial \bar{A}}{\partial c_m}(\vec{c}(X)) \frac{\partial c_m}{\partial X_i}(X)  \nonumber\\
	&&+\frac{\hbar^2}{16} \Bigg(\Bigg. \sum_{m=1}^K \sum_{i,j=1}^{2L}\frac{\Delta_{2i} \Delta_{2,j}}{2} \frac{\partial \bar{A}}{\partial c_m}(\vec{c}(X)) \frac{\partial^2 c_m}{\partial X_{1i} \partial X_{1j}}(X)\\
	&&+\sum_{m,n=1}^K \sum_{i,j=1}^L \frac{\Delta_{2i} \Delta_{2j}}{2} \frac{\partial c_m}{\partial X_{1i}}(X)\frac{\partial c_n}{\partial X_{1j}}(X) \frac{\partial^2 \bar{A}(\vec{c}(X))}{\partial c_m \partial c_n}  \Bigg.\Bigg)+ \mathcal{O}(\hbar^3) \nonumber
\end{eqnarray}
where here, as opposed to \eqref{math_OTOC2ndorder}, the $\mathcal{O}(\hbar^n)$ corrections are real $\mathcal{O}(\hbar^n)$ corrections, and not initially small errors that blow up to infinity after evolution. The same development from Section \ref{sec_SemiclassicalLimit} is carried out and we obtain again a Poisson bracket between 2 symbols, this time one of them being a phase-space average:
\begin{eqnarray} \label{otoc_longtime}
	C_\infty &=  \hbar^2 \int \dd X \left\{  \bar{A}(\vec{c}(X)), B(X)\right\}^2 W(X).
\end{eqnarray}
\indent This development is also valid for mixed phase space, with some slight limitations. The phase-space average is reduced to an average over the accessible region of phase space, which again is equal to the time average provided that the wavepacket is initially completely within the chaotic sea.

\section{Numerical simulations in Bose-Hubbard systems} \label{sec_numerics}
\begin{figure}[t]
	\centering
	\centerline{
		\includegraphics[width=\textwidth, angle=0, trim={0cm 0cm 0cm 0cm}, clip]{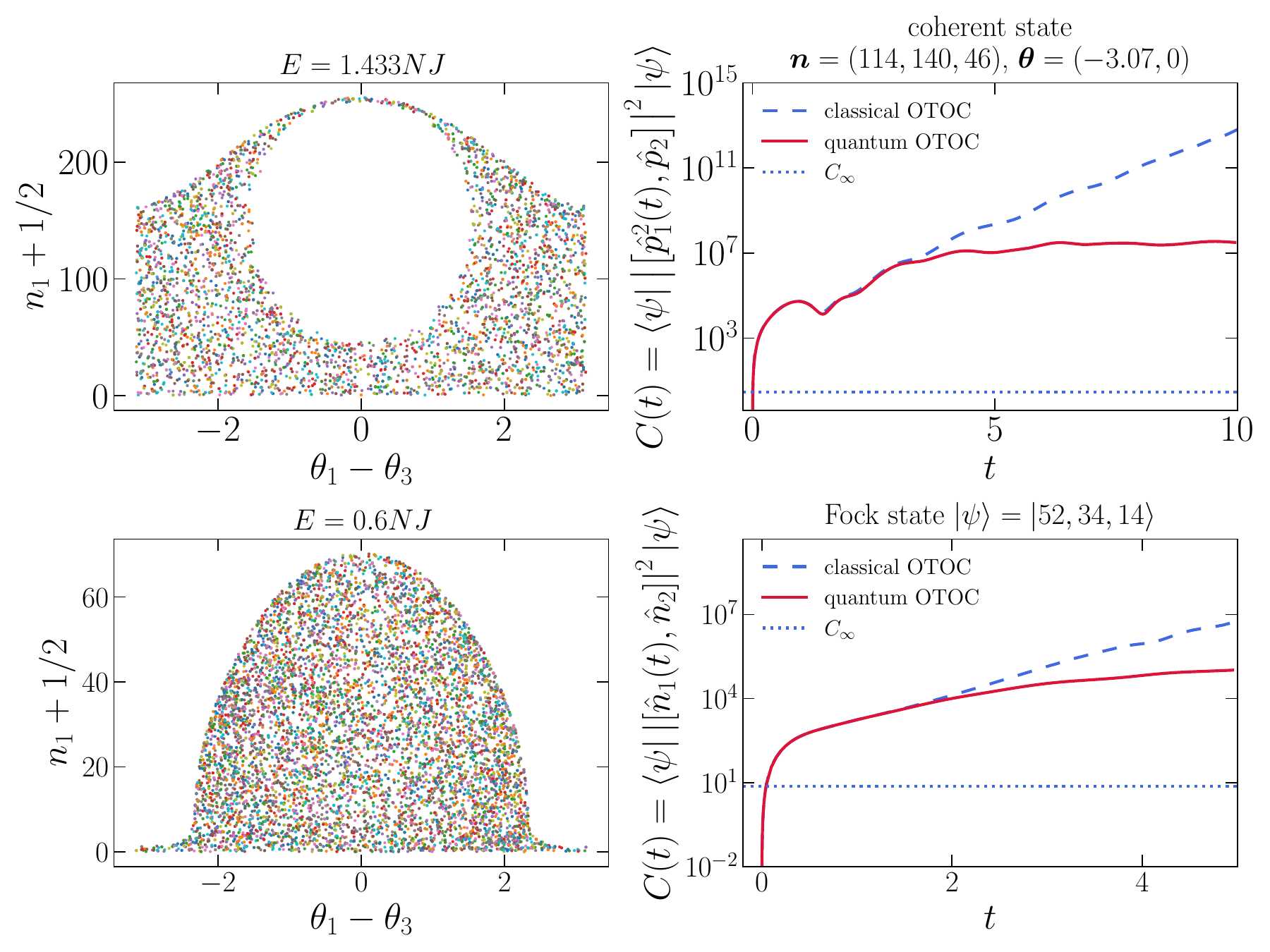}
	}
	\captionsetup{justification=justified}
	\caption{Left column: Poincaré surface of section of the Bose-Hubbard trimer for the sets of parameters (top)      $U=0.02J$, $N=300$, at energy $E=1.433NJ$ and (bottom) $U=0.06J$, $N=300$, $E=0.6NJ$. The vertical axis represents the population of the first site $n_1$ and the horizontal one the difference between the phase of the first site $\theta_1$ and the third one $\theta_3$.\\
                Right column: quantum and classical OTOC (Poisson bracket) in respectively the solid red and the dashed blue curve as well as the quasiclassical saturation value $C_\infty$ in dotted blue for (top) the coherent state centred around the occupancy levels of the Fock state $\ket{114,140,46}$ with phases $(-3.07,0,0)$, $U=0.02J$, $C_\infty = 1.18$ and (bottom) the Fock state $\ket{52, 34, 14}$, $U=0.06J$, $C_\infty = 10$.}
	\label{fig_trimer}
\end{figure}
\begin{figure}[t]
	\centerline{
		\includegraphics[width=\textwidth, angle=0, trim={0cm 0cm 0cm 0cm}, clip]{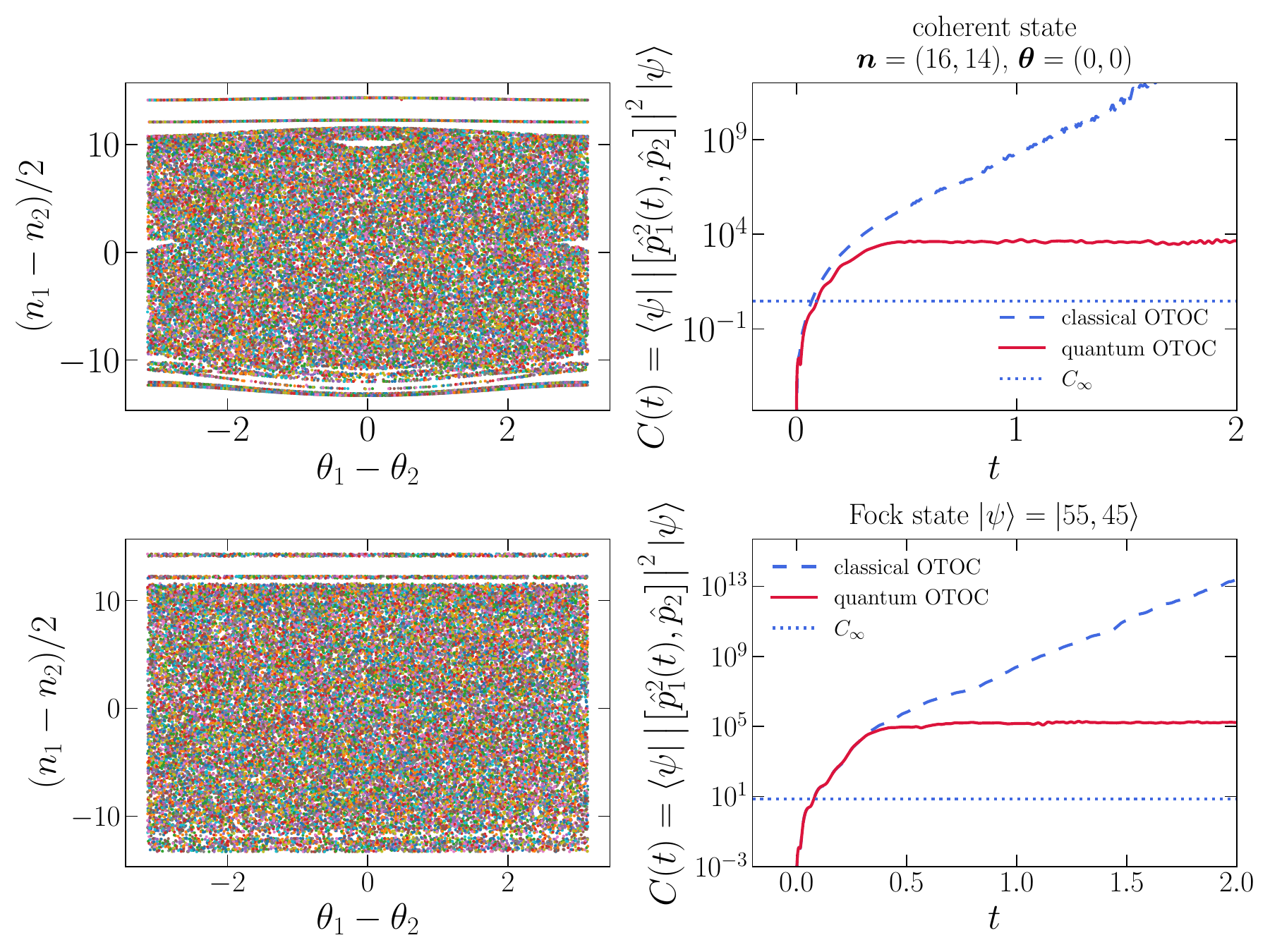}
	}
	\captionsetup{justification=justified}
	\caption{Left column: stroboscopic map of the Bose-Hubbard driven dimer for the sets of parameters (top) $U=3J$, $N=30$, $\delta = 20J$, $\omega = 10J$ and (bottom) $U=1J$, $N=100$, $\delta=20J$, $\omega = 10J$. The vertical axis represents difference of the population of the first site $n_1$ and the second one $n_2$, and the horizontal axis shows the difference between the phase of the first site $\theta_1$ and the second one $\theta_2$.\\
                Right column: quantum and classical OTOC (Poisson bracket) in respectively the solid red and the dashed blue curve as well as the quasiclassical saturation value $C_\infty$ in dotted blue for (top) the coherent state centred around the occupancy levels of the Fock state $\ket{16,14}$ with phases $(0,0)$, $U=3J$, $\delta=20J$, $\omega = 10J$, $C_\infty = 7.375$ and (bottom) the Fock state $\ket{55, 45}$, $U=1J$, $\delta=20J$, $\omega = 10J$ $C_\infty = 0.2$.}
	\label{fig_dimer}
\end{figure}
In this section, we apply our formalism to the Bose-Hubbard system. It is described by the Hamiltonian
\begin{eqnarray}
	\fl \hat{H} = \sum_{l=1}^L \left( E_l(t) \hat{n}_l  + \frac{U}{2}\ \hat{n}_l \left(\hat{n}_l -1\right) \right) - J \sum_{l=1}^{L-1} \left( \hat{b}^\dagger_l \hat{b}_{l+1}  +  \hat{b}^\dagger_{l+1} \hat{b}_l\right)
\end{eqnarray}
with $L$ the number of sites, $E_l$ is the on-site energy that can in general depend on time, $U$ the 2-body interaction strength, $J$ the hopping between adjacent site, $\hat{b}_l$ and $\hat{b}^\dagger_l$ the annihilation and creation operators, and $\hat{n}_l= \hat{b}^\dagger_l \hat{b}_l$ the population operator on site $l$ \cite{Rammensee_Urbina_Richter_2018}. We distinguish 2 cases for the on-site energy:
\begin{eqnarray}
    \begin{cases}
    L=2: E_1(t)=-E_2(t) = \delta \cos(\omega t)\\
    L=3: E_l(t)=0 \ \forall l
    \end{cases}
\end{eqnarray}
where $\delta$ and $\omega$ are the amplitude and the frequency of the driving in the case of the dimer. We first focus on the trimer without driving, and then on the driven dimer. The simulations use operators that are quadratic in terms of position and momentum operators, which means our previous derivations can be applied.\\
\indent As we expressed the  quasiclassical OTOC as a phase-space integral, a natural way to evaluate such an expression is by using the truncated Wigner method. It consists in sampling the wavepacket according to the Wigner distribution, make each sampled point evolve according to the Gross-Pitaevskii equation, and sum the contributions to evaluate the integral \cite{Dujardin_Engl_Urbina_Schlagheck_2015}.
%
%
Ideal states for investigating the quantum-classical correspondence are coherent states, since they minimise the Heisenberg uncertainty relation and have already proven to be experimentally relevant since they describe perfectly a Bose-Einstein condensate \cite{Tomsovic_Schlagheck_Ullmo_Urbina_Richter_2018}. Moreover, their Wigner function $W_{\rm CS}$ is a Gaussian \cite{Chretien_2021}, which is positive everywhere, and thus can be considered to be a probability distribution. Its expression is
\begin{eqnarray} \label{wigner_cs}
	W_{\rm{CS}}(\bm{q},\bm{p}) = \prod_{l=1}^{L}\frac{1}{\pi \hbar} \rme^{-2 (q_l - \varepsilon_{\rm{q}} \alpha_{\rm{R},l})^2/\varepsilon_{\rm{q}}^2}\ \rme^{-2 (p_l - \varepsilon_{\rm{p}} \alpha_{\rm{I},l})^2/\varepsilon_{\rm{p}}^2}
\end{eqnarray}
where the coherent state is defined by complex numbers $\left(\alpha_{\rm{R},1}+\rmi \alpha_{\rm{I},1},..., \alpha_{\rm{R},L}+\rmi \alpha_{\rm{I},L}\right)$, and $\varepsilon_{\rm{q}}, \varepsilon_{\rm{p}}$ depends on experimental parameters \cite{Chretien_2021}.
%
However, these states are not the ones that naturally emerge when working within the framework of second quantisation. Those are Fock states, or number states. Additionally, they correspond to the state obtained when preparing particles in uncoupled wells, also called Mott insulators and which correspond to the low-hopping limit of Bose-Hubbard models \cite{Gerbier_Widera_Fölling_Mandel_Gericke_Bloch_2005}. Their Wigner function contains negativities, which bear no classical meaning, but these can be considered to be strictly quantum as they disappear when looking at the classical limit \cite{Curtright_Fairlie_Zachos_2014,Kenfack_Zyczkowski_2004}. The Wigner quasidistribution of a Fock state $W_{\rm{Fock}}$ is given by \cite{Chretien_2021}
\begin{eqnarray}
	W_{\rm{Fock}}(\bm{q},\bm{p}) = \prod_{l=1}^{L} \frac{4}{\pi} (-1)^{n_l} \rme^{-2 \abs{q_l^2+ p_l^2}^2} L_{n_l} (4 \abs{q_l^2+ p_l^2}^2)
\end{eqnarray}
with $n_l$ the occupation of site $l$ and $L_n$ the $n^\text{th}$ Laguerre polynomial. It can be approximated by a Dirac delta on the occupation level in the classical limit:
\begin{eqnarray} \label{wigner_fock}
	W_{\rm{Fock},\rm{cl}}(\bm{q},\bm{p}) \to \prod_{l=1}^L\frac{1}{2\pi} \delta\left(\frac{q_l^2 + p_l^2 -1}{2} -n_l \right).
\end{eqnarray}

In the trimer, we consider $\hat{A}$ and $\hat{B}$ to be population operators acting on two sites $i$ and $j$. The corresponding classical OTOC is \cite{Richter_Urbina_Tomsovic_2022}:
\begin{eqnarray} \label{PB_2nd_CS}
	C_{\rm{cl}}(t) & = \hbar^2\int \dd \bm{n} \dd\bm{\theta} \left(\frac{\partial n_{i,t}}{\partial \theta_j}(\bm{n}, \bm{\theta}) \right)^2 W(\bm{n},\bm{\theta}).
\end{eqnarray}
and the quasiclassical long-time limit is
\begin{eqnarray} \label{PB_2nd_CS_longtime}
	C_\infty & = \hbar^2\int \dd \bm{n} \dd\bm{\theta} \left(\frac{\partial \bar{n}_{i,t}}{\partial \theta_j}(E(\bm{n}, \bm{\theta}),N( \bm{n}, \bm{\theta})) \right)^2 W( \bm{n}, \bm{\theta}).
\end{eqnarray}
where $W$ will either be $W_{\rm{CS}}$ or $W_{\rm{Fock},\rm{cl}}$.\\
\indent In the top left part of \fref{fig_trimer}, we show a Poincaré surface of section of the phase space of the trimer with the system parameters $U=0.02J$, a total of $N=300$ particles at an energy of $E=1.433J$. We represent the population of the first site $n_1$ on the vertical axis, and the phase difference between the first and last site, $\theta_1$ and $\theta_3$ modulo $2\pi$ on the horizontal axis. These sections are obtained by fixing the phase of the second site to $0$ modulo $2\pi$ and the population on the other site is fixed using the constant of motions. There are actually two solutions that fulfill these constraints, so we choose consistently the same solution.
More details on how to obtain such a section are available in \cite{Vanhaele_2016}. 
%
%
In the top right part we show the exact quantum OTOC (solid red curve) as well as the Poisson bracket (dashed blue curve) for the coherent state\footnote{In practice, we work with number-projected coherent states for the quantum simulations. However, since the experimental realisation would correspond to Bose-Einstein condensates with a fixed number of particles prepared in wells, those are more suitable.} centred around the occupancy levels of the Fock state $\ket{114,\ 140,\ 46}$, with phases $(-3.07,0,0)$, and with the same system parameters as the Poincaré section. In the quantum curve, after a brief power-law increase one can see a short exponential growth once the hyperbolic dynamics kick in, followed by a saturation after the Ehrenfest time. The classical curve matches almost perfectly the quantum one for short time. However, as it was explained in Section \ref{sec_nohbarExpansion}, obtaining the Poisson bracket was at the cost of an $\hbar$ expansion inside an evolution operator which
limits the time validity. The result is an exponential growth, exhibiting the chaotic dynamics of the system, that never stops because of the absence of quantum interferences. In addition, we plotted the long-time value as a blue horizontal dotted line. It is finite, as opposed to the Poisson bracket that growth indefinitely, but quasiclassical, \textit{i.e.} without any interferences considered. The value is however several orders of magnitude smaller than the quantum saturation one. This shows the importance of effects beyond quasiclassical physics, such as the $X$ pairing, as was stated in \cite{Rammensee_Urbina_Richter_2018} and \cite{Richter_Urbina_Tomsovic_2022}.

\indent Next, we consider a Fock state. We first represent a Poincaré section of the phase space for the system parameters $U=0.06J$ with a total population of $N=100$ at an energy $E=0.6NJ$ in the bottom left part of \fref{fig_trimer}. We also show the quantum OTOC and the Poisson bracket for the Fock state $\ket{105,100,100}$ with the same system parameters on the right. The same observations as above can be made, \textit{i.e.} an initial perfect agreement between the two curves during the power-law and the start of the exponential regime, followed by a saturation of the quantum curve whereas the classical one continues to grow. Again, the long-time value is shown as a dotted horizontal line and the conclusion is the same as with coherent states.\\
%
%
%
%
%
%
%
\indent We then focused on the driven dimer. This system has the advantage of having a phase space that can be represented in 2 dimensions only. The dynamics can thus be entirely visualised using a stroboscopic section, as shown in the left part of \fref{fig_dimer}.
%
%
%
For this system we used different operators. Because of the driving, the Hamiltonian is now time-dependent and the energy is not conserved anymore. Only the number of particles is a constant of motion, which will have consequences regarding the operators we are working with. In addition, the long-time quasiclassical OTOC can be entirely computed analytically. In \eqref{otoc_longtime}, one can apply the chain rule in the derivative of the Poisson bracket and show that in general the expression is
\begin{eqnarray}
	\left\{ \bar{A}(\vec{c}(X), B(X))\right\} = \sum_{k=1}^{K} \frac{\partial \bar{A}(\vec{c}(X))}{\partial c_k} \left\{ c_k(X), B(X)\right\}
\end{eqnarray}
where here the sum contains only one contribution since there is only one constant of motion: the total number of particles. In order to obtain non-trivial result, one has to work with operators that do not commute with the total population $\hat{N}$. We worked with momentum operators in the driven dimer, $\hat{A}=\hat{p}^2_i$ and $\hat{B} = \hat{p}_j$. We computed the OTOC and its classical limit,
\begin{eqnarray} \label{PB_2nd_CS}
	\fl C(t) = \bra{\psi} \abs{\left[\hat{p}^2_i(t), \hat{p}_j(0) \right]}^2 \ket{\psi} \qquad C_{cl}(t) = \hbar^2\int \dd  \bm{n} \dd\bm{\theta} \left(\frac{\partial p^2_{i,t}}{\partial q_j}(\bm{n}, \bm{\theta}) \right)^2 W( \bm{n}, \bm{\theta})
\end{eqnarray}
with $i=1$ and $j=2$. The long-time value of the driven dimer for this OTOC is given by
\begin{eqnarray}
	C_\infty & = \hbar^2\int \dd  \bm{n} \dd\bm{\theta} \left(\frac{\partial \bar{p^2}_{i,t}}{\partial q_j}(N(\bm{n}, \bm{\theta})) \right)^2 W(\bm{n}, \bm{\theta}).
\end{eqnarray}
The results are shown in the right part of \fref{fig_dimer} in the top half for coherent state, and in the bottom half for Fock states. We draw the same conclusions as previously.

\section{Conclusion}
In summary, we developed in this paper a full-fledged semiclassical framework for the calculation of Out-of-Time-Ordered correlators in chaotic systems via the van Vleck-Gutzwiller propagator.
By means of a diagonal approximation, this framework allowed us to yield a proper quasiclassical expression \eqref{Ccl_sumcontrib} for the OTOC, employing three different ways of trajectory pairings whose respective contributions \eqref{OTOC_pair1}--\eqref{OTOC_pair3} have to be properly added and subtracted to avoid double countings.
In the strict classical limit $\hbar \to 0$ the result of the Wigner-Moyal approach is recovered, corresponding effectively to a replacement of the quantum commutator by the classical Poisson bracket multiplied by $i\hbar$, and consequently amounting to an exponential increase $\propto e^{2\lambda t}$ of the OTOC governed by the system's Lyapunov exponent $\lambda$.

The quasiclassical expression \eqref{Ccl_sumcontrib} allowed us also to obtain a prediction for an asymptotic long-time threshold value of the OTOC.
However, comparisons with numerical results that were obtained within unperturbed and periodically driven low-dimensional Bose-Hubbard systems reveal that this threshold value grossly underestimates the actual saturation value of the quantum OTOC.
This strongly supports the conjecture that nondiagonal contributions stemming from trajectory quadruplets that exhibit low-angle crossings (i.e., those represented by the $X$ pairing in Fig.~\ref{fig_pairing}) dominantly determine the OTOC value for times larger than the system's Ehrenfest time, which would be unsurprising given the fact that for long times the phase-space weight of those trajectory quadruplets largely exceeds the weights of the trajectory pairs in the $V$ and $\Lambda$ pairings.
Owing to the similarity of the latter two, we expect that the $\Lambda$ pairing contribution, which was not evaluated for long times, amounts to a very similar value for the long-time limit of the OTOC as the $V$ pairing contribution, whereas the $I$ pairing is not expected to contribute at all in the long-time limit. 

Though exemplified only within low-dimensional Bose-Hubbard systems, the semiclassical van Vleck-Gutzwiller theory of OTOCs that we developed here is universally valid for all types of quantum single-, few- and many-body systems that exhibit a well-defined classical counterpart.
Moreover, we expect that it will also provide a valuable starting point for spin chains \cite{Pletyukhov02,Akila17,bruckmann2018} or fermionic systems \cite{fermionic_propagator,Grossmann14} for which the validity of the semiclassical approximation cannot be straightforwardly justfied.
This will open perspectives to properly refine the notion of chaos in such systems.


\appendix
\section*{Appendix}

\section{Generic quasiclassical terms} \label{sec_app_GenericTerm}
The out-of-time-ordered correlator is defined by
\begin{flalign} \label{app_OTOC_def}
C(t) &= \bra{\psi} \abs{\left[\hat{A}(t), \hat{B}(0) \right]}^2 \ket{\psi}.
\end{flalign}
When expanding the square modulus, one obtains four terms. To avoid working with long expressions from the start, let us first focus on an object of the form
\begin{flalign} \label{app_G_braket}
	O_{CDEFG}(t) &= \bra{\psi}  \hat{C} \hat{D}(t) \hat{E} \hat{F}(t) \hat{G}  \ket{\psi},
\end{flalign}
which we will call the generic quasiclassical term, and then use it to construct the OTOC by replacing $\hat{D}, \hat{F}$ by $\hat{A}$ and $\hat{C}, \hat{E}, \hat{G}$ by either $\mathds{1}$, $\hat{B}$ or $\hat{B}^2$ depending on the term of the OTOC:
\begin{flalign} \label{eq_app_OTOCFromGeneric}
    C(t) &= O_{BA \mathds{1} AB}(t) - O_{BABA\mathds{1}}(t) - O_{\mathds{1}ABAB}(t) + O_{\mathds{1}AB^2 A\mathds{1}}(t). 
\end{flalign} 

\subsection{Full semiclassical expression} \label{sec_app_FullSemiclassicalExpression}
By introducing completeness relations of position eigenstates $\hat{1} =\int d\bm{x} \ket{\bm{x}} \bra{\bm{x}}$ between each operators, one can reexpress the expectation value \eqref{app_G_braket} as a phase-space integral of matrix elements and propagators:
\begin{flalign} \label{app_G}
   O_{CDEFG}(t) &=\int \prod_{i=1}^6 \dd \bm{x}_i \prod_{j=2}^5 \dd \bm{y}_j \ K^* (\bm{y}_2, \bm{x}_2,t) K(\bm{y}_3, \bm{x}_3,t) K^*(\bm{y}_4, \bm{x}_4,t) K(\bm{y}_5, \bm{x}_5,t) \nonumber& \\
    &\quad \times\matrixel{\bm{x}_1}{\hat{C}}{\bm{x}_2}\matrixel{\bm{y}_2}{\hat{D}}{\bm{y}_3}\matrixel{\bm{x}_3}{\hat{E}}{\bm{x}_4}\matrixel{\bm{y}_4}{\hat{F}}{\bm{y}_5}\matrixel{\bm{x}_5}{\hat{G}}{\bm{x}_6}&
\end{flalign} 
with \mbox{$K(\bm{q}^{\rm{f}}, \bm{q}^{\rm{i}}, t) = \bra{\bm{q}^{\rm{f}}} \hat{U}(t) \ket{\bm{q}^{\rm{i}}}$} the propagator from $\bm{q}^{\rm{i}}$ to $\bm{q}^{\rm{f}}$ in time $t$ \cite{Stockmann_1999}\cite{Rammensee_2019}. \\
\indent We then recall here the expression of the Weyl symbol of an operator $\hat{O}$:
\begin{eqnarray} \label{app_def_weylSymbol}
    O(\bm{q},\bm{p}) = \int \dd \bm{\xi} \bra{\bm{q}+\frac{\bm{\xi}}{2}} \hat{O} \ket{\bm{q}-\frac{\bm{\xi}}{2}} \rme^{-\rmi \bm{p}\cdot \bm{\xi}/\hbar}
\end{eqnarray}
of the Wigner function of a pure state $\ket{\psi}$:
\begin{eqnarray}
	W(\bm{q}, \bm{p}) = \frac{1}{(2\pi \hbar)^L}\int \dd \bm{\xi}\ \psi^*\left( \bm{q} + \frac{\bm{\xi}}{2} \right) \psi\left( \bm{q} - \frac{\bm{\xi}}{2} \right) \rme^{\rmi\bm{p}\cdot \bm{\xi}/\hbar},
\end{eqnarray}
and of the semiclassical van Vleck-Gutzwiller propagator:
\begin{eqnarray}
    K(\bm{q}^{\rm{f}}, \bm{q}^{\rm{i}}, t) & \simeq \sum_{\gamma:\bm{q}^{\rm{i}}\to \bm{q}^{\rm{f}}} A_\gamma (\bm{q}^{\rm{f}}, \bm{q}^{\rm{i}}, t)\ \rme^{\rmi R_\gamma(\bm{q}^{\rm{f}}, \bm{q}^{\rm{i}}, t)/\hbar}
\end{eqnarray}
with
\begin{eqnarray}
    A_\gamma (\bm{q}^{\rm{f}}, \bm{q}^{\rm{i}}, t) &= \frac{1}{(2\pi \hbar)^{L/2}} \abs{\det \frac{\partial^2 R_\gamma}{\partial \bm{q}^f \partial \bm{q}^{\rm{i}}} (\bm{q}^{\rm{f}},\bm{q}^{\rm{i}},t)}^{1/2} \rme^{- i \mu_\gamma \pi/4},\\
    R_\gamma(\bm{q}^{\rm{f}}, \bm{q}^{\rm{i}}, t) &= \int_{0}^{t} \dd t^\prime  \left( \bm{p}_\gamma(t^\prime) \cdot \bm{\dot{q}}_\gamma (t) - H_{\rm{cl}} (\bm{q}_\gamma (t), \bm{p}_\gamma (t))\right) 
\end{eqnarray}
respectively the amplitude and the action of the trajectory, with $\gamma$ the family of trajectories, $\nu_\gamma$ Maslov's index and $\bm{q}_\gamma$ and $\bm{p}_\gamma$ the position and momentum related to the trajectory $\gamma$ \cite{Rammensee_Urbina_Richter_2018}\cite{Engl_Urbina_Richter_2016}. Injecting these in \eqref{app_G}, we obtain:
\begin{flalign} \label{app_G_vvG}
	 O_{CDEFG}(t) =\ &\frac{1}{(2\pi \hbar)^{5L}} \int \prod_{i=1}^6 \dd\bm{x}_i \int\prod_{j=2}^5 \dd \bm{y}_j \dd \bm{P}\ W\left(\frac{\bm{x}_1 + \bm{x}_6}{2}, \bm{P}\right)& \nonumber\\
    & \times \sum_{\substack{\alpha:\bm{x}_2\to \bm{y}_2 \\\beta:\bm{x}_3\to \bm{y}_3}} \sum_ {\substack{\gamma:\bm{x}_4\to \bm{y}_4\\ \delta:\bm{x}_5\to \bm{y}_5}} A_\alpha^* A_\beta A_\gamma^* A_\delta& \nonumber \\ 
	&\times  \exp\left(\frac{\rmi}{\hbar} \left[ R_\beta(\bm{y}_3,\bm{x}_3,t) - R_\alpha(\bm{y}_2,\bm{x}_2,t) - R_\gamma(\bm{y}_4,\bm{x}_4,t) + R_\delta(\bm{y}_5,\bm{x}_5,t)\right] \right) & \nonumber\\
    & \times \int \prod_{k=1}^5 \dd\bm{p}_k\ C\left(\frac{\bm{x}_1 + \bm{x}_2}{2}, \bm{p}_1 \right) D\left(\frac{\bm{y}_2 + \bm{y}_3}{2}, \bm{p}_2\right)& \nonumber\\
	& \times  E\left(\frac{\bm{x}_3 + \bm{x}_4}{2}, \bm{p}_3\right)  F\left(\frac{\bm{y}_4 + \bm{y}_5}{2}, \bm{p}_4\right) G\left(\frac{\bm{x}_5 + \bm{x}_6}{2}, \bm{p}_5\right)& \nonumber \\ 
	&\times \exp\left[\rmi ( \bm{p}_1 \cdot (\bm{x}_1-\bm{x}_2) + \bm{p}_2 \cdot (\bm{y}_2-\bm{y}_3) + \bm{p}_3 \cdot (\bm{x}_3-\bm{x}_4))/\hbar\right] & \nonumber\\
 & \times \exp\left[ \rmi ( \bm{p}_4 \cdot (\bm{y}_4-\bm{y}_5) +  \bm{p}_5 \cdot (\bm{x}_5-\bm{x}_6) -\bm{P}\cdot (\bm{x}_1 - \bm{x}_6))/ \hbar\right]&
\end{flalign}
using the shorthand notation $\sum_{\alpha:\bm{x}\to \bm{y}} A_\alpha \equiv \sum_{\zeta:\bm{x}\to \bm{y}} A_\alpha(\bm{y},\bm{x},t)$. This expression is the starting point of the derivations that follow.\\
\indent The next step is to use the diagonal approximation. This results in four contributions and are shown in \ref{fig_pairing}. In practice, the last one is related to phenomena beyond quasiclassics and thus does not contribute. We can then write
\begin{eqnarray}
    G(t) = G^V(t) + G^\Lambda(t) - G^I(t)
\end{eqnarray}
where the minus is there to avoid overcounting the $I$ pairing.

\subsection{$V$ pairing}
The first pairing corresponds to the situation where $\alpha \sim \beta$ and $\gamma \sim \delta$, meaning that $\alpha$ and $\beta$ are actually the same family, likewise for $\gamma$ and $\delta$. The first step is to exploit the fact that when two trajectories belong to the same family, they can be linearised around a central one. We thus switch to some centre-of-mass and relative coordinates:
\begin{eqnarray}
	 \bm{x}_1 &= \bm{\mathcal{X}} + \frac{\hbar\bm{x}}{2} \qquad \bm{x}_2 &= \bm{R}_2 + \frac{\hbar \bm{r}_2}{2} \qquad \bm{x}_4 = \bm{R}_4+ \frac{\hbar \bm{r}_4}{2} \nonumber\\
	 \bm{x}_6 &= \bm{\mathcal{X}} - \frac{\hbar \bm{x}}{2} \qquad \bm{x}_3 &= \bm{R}_2 - \frac{\hbar \bm{r}_2}{2} \qquad \bm{x}_5 = \bm{R}_4 - \frac{\hbar \bm{r}_4}{2} \nonumber\\
      \bm{y}_2 &= \bm{S}_2 + \frac{\hbar \bm{s}_2}{2} \qquad  \bm{y}_4 &= \bm{S}_4 + \frac{\hbar \bm{s}_4}{2} \nonumber\\
      \bm{y}_3 &= \bm{S}_2 - \frac{\hbar \bm{s}_2}{2} \qquad  \bm{y}_5 &= \bm{S}_4 - \frac{\hbar \bm{s}_4}{2}.
\end{eqnarray}
The amplitudes of the propagator are considered at the $0^\text{th}$ order only. More specifically
\begin{eqnarray}
    A^*_\alpha (\bm{S}+\frac{\hbar \bm{s}}{2}, \bm{R} + \frac{\hbar \bm{r}}{2}) A_\alpha (\bm{S}-\frac{\hbar \bm{s}}{2}, \bm{R} - \frac{\hbar \bm{r}}{2}) \simeq\abs{A_\alpha(\bm{S}, \bm{R})}^2
\end{eqnarray}
up to correction that scale as $\hbar^2$.
The first non-vanishing terms when expanding the actions in the phase are of the $1^\text{st}$ order:
\begin{flalign}
    R_\alpha \left(\bm{S}_2 + \frac{\hbar \bm{s}_2}{2}, \bm{R}_2 + \frac{\hbar \bm{r}_2}{2} ,t\right) &- R_\alpha \left(\bm{S}_2 - \frac{\hbar \bm{s}_2}{2}, \bm{R}_2 - \frac{\hbar \bm{r}_2}{2} ,t\right) & \nonumber\\
    &\simeq \hbar \bm{s}_2 \cdot\bm{p}_\alpha ^{\rm{f}}(\bm{S}_2,\bm{R}_2,t) - \hbar \bm{r}_2 \cdot\bm{p}_\alpha ^{\rm{i}}(\bm{S}_2,\bm{R}_2,t)&
\end{flalign}
where we have used $\partial R_\gamma/\partial \bm{q}^{\rm{i}}(\bm{q}^{\rm{f}}, \bm{q}^{\rm{i}},t) = - \bm{p}_\gamma ^{\rm{i}}(\bm{q}^{\rm{f}}, \bm{q}^{\rm{i}},t)$ and $\partial R_\gamma/\partial \bm{q}^{\rm{f}}(\bm{q}^{\rm{f}}, \bm{q}^{\rm{i}},t) = \bm{p}_\gamma ^{\rm{f}}(\bm{q}^{\rm{f}}, \bm{q}^{\rm{i}},t)$.  This yields 
\begin{flalign} 
	 O_{CDEFG}^V(t) =\ &\frac{1}{(2\pi)^{5L}} \int \dd \bm{R}_{2,4} \dd \bm{r}_{2,4} \dd \bm{S}_{2,4} \dd \bm{s}_{2,4} \dd \bm{\mathcal{X}} \dd \bm{x}\ \sum_
	{\substack{\alpha:\bm{R}_2\to \bm{S}_2  \\ \gamma:\bm{R}_4\to \bm{S}_4}}
	\abs{A_\alpha}^2 \abs{A_\gamma}^2\ W\left(\bm{\mathcal{X}}, \bm{P}\right)& \nonumber \\ 
    &\times  \int \prod_{i=1}^5 \dd \bm{p}_i\ C\left(\frac{\bm{\mathcal{X}} + \textbf{R}_2}{2} +\frac{\hbar(\bm{x}+ \bm{r}_2)}{4}, \bm{p}_1\right)  D\left(\bm{S}_2 , \bm{p}_2\right)& \nonumber \\
	&\times  E\left(\frac{\bm{R}_2 + \textbf{R}_4}{2} +\frac{\hbar(\bm{r}_4 - \bm{r}_2)}{4}, \bm{p}_3\right) \nonumber F\left(\bm{S}_4 , \bm{p}_4\right)& \nonumber\\
 &\times G\left(\frac{\bm{\mathcal{X}} + \textbf{R}_4}{2} -\frac{\hbar(\bm{x}+ \bm{r}_4)}{4}, \bm{p}_5 \right) \exp[\rmi \left(\bm{s}_2\cdot(\bm{p}_2- \bm{p}_\alpha^{\rm{f}})  \right)]& \nonumber\\
    &\times \exp[\rmi \left(\bm{r}_2\cdot(\bm{p}_\alpha^{\rm{i}} - \tfrac{\bm{p}_1+\bm{p}_3}{2}) +\bm{s}_4\cdot(\bm{p}_4-\bm{p}_\gamma^{\rm{f}}) + \bm{r}_4\cdot(\bm{p}_\gamma^{\rm{i}} - \tfrac{\bm{p}_3 + \bm{p}_5}{2}) \right)]&\nonumber\\
    &\times \exp[\rmi \left(\bm{\mathcal{X}}\cdot(\bm{p}_1 - \bm{p}_5) +\bm{x}\cdot \tfrac{\bm{p}_1 + \bm{p}_5}{2}  + \bm{R}_2\cdot(\bm{p}_3 - \textbf{p}_1 ) + \bm{R}_4\cdot(\bm{p}_5-\bm{p}_3) \right)]& \nonumber\\
    &\times \exp[-\rmi \bm{P}\cdot\bm{x}]&.
\end{flalign}
\indent In order to make the last step towards the symplectic formalism, we will introduce a symmetry between the position and momentum arguments of a given symbol, and between two pairs of symbols as well. To this end, we will perform several changes of variables to some new centre-of-mass and relative coordinates. First, we do it for the position argument of symbols $C$ and $G$:
\begin{flalign}
    \bm{Y} &= \frac{1}{2} \left( \frac{\bm{\mathcal{X}} + \textbf{R}_2}{2} +\frac{\hbar(\bm{x}+ \bm{r}_2)}{4} + \frac{\bm{\mathcal{X}} + \textbf{R}_4}{2} +\frac{\hbar(\bm{x}+ \bm{r}_4)}{4} \right) \qquad \bm{p}_1 = \bm{\Pi}_1 + \frac{\hbar \bm{\omega}_1}{2}& \nonumber \\
    \bm{y} &= \frac{1}{2} \left( \frac{\bm{\mathcal{X}} + \textbf{R}_2}{2} +\frac{\hbar(\bm{x}+ \bm{r}_2)}{4} - \frac{\bm{\mathcal{X}} + \textbf{R}_4}{2} -\frac{\hbar(\bm{x}+ \bm{r}_4)}{4} \right) \qquad \bm{p}_5 = \bm{\Pi}_1 - \frac{\hbar\bm{\omega}_1}{2}.&
\end{flalign}
We then use the amplitudes of the propagator as a Jacobian of a change of variables to change the integration with respect to the final positions $\bm{S}_2, \bm{S}_4$ to initial momenta $\bm{p}_\alpha^i, \bm{p}_\beta^i$:
\begin{flalign}
	\sum_\alpha \abs{A_\alpha}^2 &= \sum_\alpha \frac{1}{(2\pi \hbar)^{L}}\abs{\det \frac{\partial^2 R}{\partial \bm{S}_2 \partial \bm{R}_2} (\bm{S}_2,\bm{R}_2)} = \sum_\alpha \frac{1}{(2\pi \hbar)^{L}}\abs{\frac{\partial \bm{p}_\alpha^{\rm{i}}}{\partial \bm{S}_2}}&
\end{flalign}
where we have used this time 
\mbox{$\partial R_\gamma/\partial \bm{q}^{\rm{f}}(\bm{q}^{\rm{f}}, \bm{q}^{\rm{i}},t) = \bm{p}_\gamma ^{\rm{f}}(\bm{q}^{\rm{f}}, \bm{q}^{\rm{i}},t)$}.
We then go to centre-of-mass and relative coordinate for the symbols $D$ and $F$:
\begin{eqnarray}
	\bm{R}_2 &= \bm{R} + \frac{\hbar\bm{r}}{2} \qquad \bm{p}_\alpha^{\rm{i}} &= \bm{\Pi}_2 + \frac{\hbar\bm{\omega}_2}{2} \nonumber\\
    \bm{R}_4 &= \bm{R} - \frac{\hbar\bm{r}}{2} \qquad \bm{p}_\beta^{\rm{i}} &= \bm{\Pi}_2 - \frac{\hbar \bm{\omega}_2}{2}.
\end{eqnarray}
We perform the integral with respect to $\bm{x}$ and obtain
\begin{flalign}
    O_{CDEFG}^V(t) =\ &\frac{1}{(2\pi)^{4L}} \int \dd\bm{\mathcal{X}} \dd\bm{x} \dd\bm{Y} \dd\bm{y} \dd\bm{R} \dd\bm{r} \dd\bm{P} \dd \bm{\Pi}_{1,2} \dd\bm{\omega}_{1,2}\  \exp\left(\rmi(  \bm{\omega}_2\cdot (2 \bm{Y} - \bm{\mathcal{X}} - \bm{R}) )\right) &\nonumber\\
   &\times \exp\left(\rmi( \bm{\omega}_1 \cdot(\bm{\mathcal{X}} - \bm{R})  + \bm{y}\cdot(\bm{\Pi}_1 - \bm{P})	+ \bm{r}\cdot ( \bm{P} + \bm{\Pi}_2 - 2 \bm{\Pi}_1)) \right) & \nonumber \\
    & \times C\left(\bm{Y} +\frac{\hbar\bm{y}}{4}, \bm{\Pi}_1 + \frac{\hbar \bm{\omega}_1}{4} \right)  D\left( \left(\bm{R}_t, \bm{P}_t \right) \left(\bm{R} +\frac{\hbar\bm{r}}{4}, \bm{\Pi}_2 + \frac{\hbar \bm{\omega}_2}{4} \right) \right) &\nonumber \\
    &\times E\left(\bm{\mathcal{X}} - 2\bm{Y} + 2 \bm{R}, \bm{P} - 2 \bm{\Pi}_1 + 2 \bm{\Pi}_2 \right) F\left( \left(\bm{R}_t, \bm{P}_t \right) \left(\bm{R} -\frac{\hbar\bm{r}}{4}, \bm{\Pi}_2 - \frac{\hbar \bm{\omega}_2}{4} \right) \right) &\nonumber \\ 
    &\times G\left(\bm{Y} -\frac{\hbar\bm{y}}{4}, \bm{\Pi}_1 - \frac{\hbar \bm{\omega}_1}{4}  \right) W(\bm{\mathcal{X}}, \bm{P})&
\end{flalign}
where we used the notation $(\bm{R}_t, \bm{P}_t ) (\bm{R}, \bm{\Pi}) = (\bm{R}_t(\bm{R}, \bm{\Pi}) , \bm{P}_t(\bm{R}, \bm{\Pi})) $ to shorten the expression, and where $\bm{R}_t\left(\bm{R}, \bm{\Pi}\right)$ and $\bm{P}_t\left(\bm{R}, \bm{\Pi}\right)$ means respectively the position and momentum after an evolution for $t$ time, having started from $\left(\bm{R}, \bm{\Pi}\right)$.\\
\indent In the expression for all the pairings there is eventually a symmetry between the position and momentum argument which allows us to regroup them in a single vector, called a symplectic vector. In the case of the $V$ pairing, we can write
\begin{flalign} \label{app_G_pair1_sympl}
    O_{CDEFG}^V(t) =\ &\frac{1}{(2\pi)^{4L}} \int \dd X_{1,2,3} \dd\Delta_{1,2}\ \rme^{\rmi\Delta_1 \wedge (X_2 - X_3)} \rme^{\rmi \Delta_2 \wedge (X_1 - 2X_2 + X_3) } & \nonumber \\
    & \times  C\left(X_2 +  \frac{\hbar\Delta_1}{4}\right) D\left(X_t\left(X_1 + \frac{\hbar \Delta_2}{4}\right)\right) E(2 X_1 -2 X_2 + X_3) &\nonumber\\
    & \times F\left(X_t \left(X_1 - \frac{\hbar \Delta_2}{4}\right)\right)  G\left(X_2 -  \frac{\hbar\Delta_1}{4}\right) W(X_3)&
\end{flalign}
where $X_1 = (\bm{Y}, \bm{\Pi}_1)$, $X_2 = (\bm{R}, \bm{\Pi_2})$, $X_3 = (\bm{\mathcal{X}}, \bm{P})$, $\Delta_1 = (\bm{y}, \bm{\omega}_1)$, $\Delta_2 = (\bm{r}, \bm{\omega}_2)$, and $X_t$ indicates the evolution to time $t$.\\

\subsection{$\Lambda$ pairing}
The second pairing follows the same steps as the first one, although with some changes when it comes to the integration variables. The main difference is that one naturally ends up with integrals with respect to final conditions instead of initial ones.\\
\indent Starting from \eqref{app_G_vvG}, this time the pair of trajectories are $\alpha \sim \delta$ and $\beta \sim \gamma$. Similarly to the $V$ pairing, we go to some new centre-of-mass and relative coordinates: 
\begin{eqnarray}
	\bm{x}_1 &= \bm{\mathcal{X}} + \frac{\hbar\bm{x}}{2} \qquad \bm{x}_2 &= \bm{R}_2 + \frac{\hbar\bm{r}_2}{2} \qquad \bm{x}_3 = \bm{R}_3+ \frac{\hbar\bm{r}_3}{2}  \nonumber\\
	\bm{x}_6 &= \bm{\mathcal{X}} - \frac{\hbar\bm{x}}{2} \qquad \bm{x}_5 &= \bm{R}_2 - \frac{\hbar\bm{r}_2}{2} \qquad \bm{x}_4 = \bm{R}_3 - \frac{\hbar\bm{r}_3}{2}\nonumber\\
    \bm{y}_2 &= \bm{S}_2 + \frac{\hbar\bm{s}_2}{2} \qquad  \bm{y}_3 &= \bm{S}_3 + \frac{\hbar\bm{s}_3}{2} \nonumber\\
    \bm{y}_5 &= \bm{S}_2 - \frac{\hbar\bm{s}_2}{2} \qquad  \bm{y}_4 &= \bm{S}_3 - \frac{\hbar\bm{s}_3}{2} .
\end{eqnarray}
We can then linearise the action around these new centre-of-mass coordinates, which calls for some other changes of variables, this time related to the momenta:
\begin{eqnarray}
	\bm{p}_1 &= \bm{\Pi}_1 + \frac{\hbar\bm{\omega}_1}{2} \qquad \bm{p}_2 &= \bm{\Pi}_2 + \frac{\hbar\bm{\omega}_2}{2} \nonumber\\
	\bm{p}_5 &= \bm{\Pi}_1 - \frac{\hbar\bm{\omega}_1}{2} \qquad \bm{p}_4 &= \bm{\Pi}_2 - \frac{\hbar\bm{\omega}_2}{2}.
\end{eqnarray} 
By performing the integration with respect to $\bm{r}_3$ one can rewrite \eqref{app_G_vvG} as
\begin{flalign}
    O_{CDEFG}^\Lambda(t) =\ &\int \dd \bm{\mathcal{X}}  \dd \bm{\Delta\mathcal{X}} \dd \bm{R}_{2,3}\dd \bm{r}_{2}\dd\bm{S}_{2,3}\dd\bm{s}_{2,3}\dd\bm{\Pi}_{1,2}\dd\bm{\omega}_{1,2}\ W\left(\bm{\mathcal{X}}, \bm{P}\right)&  \nonumber\\
	&\times \sum_{\alpha,\beta} \frac{\abs{A_\alpha}^2 \abs{A_\beta}^2}{(2\pi)^{4L}}\ \exp\left[\rmi \left(\bm{s}_2 \cdot(\bm{\Pi_2} - \bm{p}_\alpha^{\rm{f}}) + \bm{r}_2\cdot(\bm{p}_\alpha^{\rm{i}} - \bm{\Pi}_1)\right)\right] &\nonumber\\
 &\times \exp\left[\rmi\left(\bm{s}_3\cdot(\bm{p}_\beta^{\rm{f}} - \Pi_2) + (\bm{\mathcal{X}} - \bm{R}_2)\cdot \bm{\omega}_1 +(\bm{S}_2-\bm{S}_3)\cdot\bm{\omega}_2 -\bm{P}\cdot\bm{x} \right) \right] &\nonumber \\
	&\times C\left(\frac{\bm{\mathcal{X}} + \bm{R}_2}{2} + \frac{\hbar(\bm{\Delta\mathcal{X}} + \bm{r}_2)}{4}, \bm{\Pi}_1 + \frac{\hbar\bm{\omega}_1}{2}\right) &\nonumber\\
 &\times D\left(\frac{\bm{S}_2 + \bm{S}_3}{2} + \frac{\hbar(\bm{s}_2 + \bm{s}_3)}{4}, \bm{\Pi}_2 + \frac{\hbar\bm{\omega}_2}{2}\right) E\left(\bm{R}_3, \bm{p}_\beta^{\rm{i}}\right) &\nonumber \\ 
	& \times F\left(\frac{\bm{S}_2 + \bm{S}_3}{2} - \frac{\hbar(\bm{s}_2 + \bm{s}_3)}{4}, \bm{\Pi}_2 - \frac{\hbar\bm{\omega}_2}{2}\right) &\nonumber\\
    &\times G\left(\frac{\bm{\mathcal{X}} + \bm{R}_2}{2} - \frac{\hbar(\bm{\Delta\mathcal{X}} + \bm{r}_2)}{4}, \bm{\Pi}_1 - \frac{\hbar\bm{\omega}_1}{2}\right). &
\end{flalign}
In order to obtain an expression that is writeable in symplectic formalism, one must perform additional change of variables that consists of shifts to symmetrise the position and momentum arguments of a given symbol. Again, one uses the amplitudes of the propagator as a Jacobian of a change of variables, this time from initial position $\bm{R}_2, \bm{R}_3$ to final momenta $\bm{p}_\alpha^{\rm{f}}, \bm{p}_\beta^{\rm{f}}$:
\begin{eqnarray} \label{app_jacobian_Rtop}
	\sum_\alpha \abs{A_\alpha}^2 = \sum_\alpha \frac{1}{(2\pi \hbar)^{L}}\abs{\frac{\partial \bm{p}_\alpha^{\rm{f}}}{\partial \bm{R}_2}}
\end{eqnarray}
where we have used \mbox{$\partial R_\gamma/\partial \bm{q}^{\rm{f}}(\bm{q}^{\rm{f}}, \bm{q}^{\rm{i}},t) = \bm{p}_\gamma ^{\rm{f}}(\bm{q}^{\rm{f}}, \bm{q}^{\rm{i}},t)$} this time. The final expression is then
\begin{flalign}  
    O_{CDEFG}^\Lambda(t) =\ &\frac{1}{(2\pi)^{4L}} \int \dd \bm{\mathcal{X}}  \dd\bm{r}_{2} \dd\bm{S}_{2,3} \dd\bm{s}_{2} \dd\bm{p}_{\alpha,\beta}^{\rm{f}} \dd\bm{\Pi}_{1,2} \dd\bm{\omega}_{1,2}\ \exp\left[] \rmi \left(\bm{s}_2\cdot (\bm{p}_\beta^{\rm{f}} - \bm{p}_\alpha^{\rm{f}})\right)\right] &\nonumber\\
    &\times \exp\left[\rmi \left( \bm{r}_2\cdot(\bm{P}_{-t}(2\bm{S}_2 - \bm{S}_3, 2\bm{p}^{\rm{f}}_\alpha - \bm{p}^{\rm{f}}_\beta)  - \bm{\Pi}_1)\right) \right] &\nonumber \\
    &\times \exp\left[] \rmi \left( (\bm{\mathcal{X}} - \bm{R}_{-t}(2\bm{S}_2 - \bm{S}_3, 2\bm{p}^{\rm{f}}_\alpha - \bm{p}^{\rm{f}}_\beta) \cdot \bm{\omega}_1 +(\bm{S}_2-\bm{S}_3)\cdot\bm{\omega}_2 \right) \right] &\nonumber\\
    &\times C\left(\bm{\mathcal{X}} +  \frac{\hbar\bm{r}_2}{4}, \bm{\Pi}_1 + \frac{\hbar\bm{\omega}_1}{4}\right) D\left(\bm{S}_2+ \frac{\hbar \bm{s}_2}{4}, \bm{p}_\alpha^{\rm{f}} + \frac{\hbar \bm{\omega}_2}{4}\right)& \nonumber\\
    &\times  E\left(\bm{R}_{-t}(\bm{S}_3, \bm{p}_\beta^{\rm{f}}), \bm{P}_{-t}(\bm{S}_3, \bm{p}_\beta^{\rm{f}})\right) F\left(\bm{S}_2 - \frac{\hbar \bm{s}_2}{4}, \bm{p}_\alpha^{\rm{f}} - \frac{\hbar\bm{\omega}_2}{4}\right) &\nonumber\\
    &\times G\left(\bm{\mathcal{X}} - \frac{\hbar \bm{\bm{r}}_2}{4}, \bm{\Pi}_1 - \frac{\hbar \bm{\omega}_1}{4}\right) &\\
    &\times W\left(2\bm{\mathcal{X}} - \bm{R}_{-t}(2\bm{S}_2 - \bm{S}_3, 2\bm{p}_\alpha^{\rm{f}} -\bm{p}_\beta^{\rm{f}}), 2\bm{\Pi}_1 - \bm{P}_{-t}(2\bm{S}_2 -\bm{S}_3, 2\bm{p}_\alpha^{\rm{f}} -\bm{p}_\beta^{\rm{f}})\right) &\nonumber
\end{flalign}
where $\bm{R}_{-t}(\bm{S}, \bm{P})$ and $\bm{P}_{-t}(\bm{S}, \bm{P})$ are the position and momentum obtained from the backward evolution of $(\bm{S}, \bm{P})$ for time $t$.\\
\indent The symplectic notation reads
\begin{flalign}\label{app_G_pair2_sympl}
    O_{CDEFG}^\Lambda(t) =\ &\frac{1}{(2\pi)^{4L}} \int \dd X \dd S_{2,3} \dd \Delta_{2,3}\ \rme^{\rmi \Delta_1\wedge(X-X_{-t}(2S_2-S_3))} \rme^{\rmi \Delta_2\wedge (S_3 - S_2)} &\nonumber \\
	& \times C\left(X+\frac{\hbar \Delta_1}{4}\right) D\left(S_2+\frac{\hbar \Delta_2}{4}\right)  E\left(X_{-t}(S_3)\right)  F\left(S_2-\frac{\hbar \Delta_2}{4}\right) & \nonumber\\
 &\times G\left(X-\frac{\hbar \Delta_1}{4}\right) W(2X-X_{-t}(2S_2-S_3))&
\end{flalign}
where $X = (\bm{\mathcal{X}}, \bm{\Pi}_1)$, $S_2 = (\bm{S}_2, \bm{p}^{\rm{f}}_\alpha)$, $S_3 = (\bm{S}_3, \bm{p}^{\rm{f}}_\beta)$, $\Delta_1 = (\bm{r}_2, \bm{\omega}_1)$, $\Delta_2 = (\bm{s}_2, \bm{\omega}_2)$, and $X_{-t}(S)$ is the backward evolution from $S$ for time $t$.\\

\subsection{$I$ pairing}
When summing the contributions from both the $V$ and $\Lambda$ pairing, we are overcounting the subcase where all four trajectories are near one another, \textit{i.e.} $\alpha \sim \beta \sim \gamma \sim \delta$. Therefore, we actually have to subtract this contribution to account for it only once.
Intuitively, one would try to linearise all four trajectories around a global central one. However, using the same set of variables as for the $\Lambda$ pairing seems far more suitable for the derivation. One can thus follow the same step as for the last pairing and obtain
\begin{flalign} \label{app_G_pair3}
    O_{CDEFG}^I(t) =\ & \frac{1}{(2\pi)^{4L} 2^{2L}} \int \dd \bm{\mathcal{X}} \dd \bm{R}_{2,3} \dd\bm{r}_{2} \dd\bm{S}_{2,3} \dd\bm{s}_{2} \dd\bm{\Pi}_{1,2} \dd\bm{\omega}_{1,2} \sum_\alpha \abs{\det\frac{\partial^2 R_\alpha(\bm{S}_2,\bm{R}_2)}{\partial \bm{S}_2 \partial \bm{R}_2}} & \nonumber\\
	&\times \abs{\det\frac{\partial^2 R_\alpha(\bm{S}_3,\bm{R}_3)}{\partial \bm{S}_3 \partial \bm{R}_3}} \exp\left( \rmi (\bm{s}_2 \cdot(\bm{\Pi}_2 -\bm{p}_\alpha^{\rm{f}}(\bm{S}_2,\bm{R}_2)) ) \right) &\nonumber\\
    & \times \exp[\left(\bm{r}_2\cdot(\bm{p}_\alpha^{\rm{i}}(\bm{S}_2,\bm{R}_2) - \bm{\Pi}_1) + (\bm{\mathcal{X}} - \bm{R}_2)\cdot \bm{\omega}_1 +\frac{\bm{S}_2-\bm{S}_3}{2}\cdot\bm{\omega}_2  \right)]&\nonumber \\
	&\times  C\left(\bm{\mathcal{X}}  + \frac{\hbar \bm{r}_2}{4}, \bm{\Pi}_1 + \frac{\hbar \bm{\omega}_1}{4}\right) & \nonumber\\
        &\times D\left(\frac{\bm{S}_2+\bm{S}_3}{2}+ \frac{\hbar \bm{s}_2}{4}, \frac{\bm{p}_\alpha^{\rm{f}}(\bm{S}_2,\bm{R}_2) + \bm{p}_\alpha^{\rm{f}}(\bm{S}_3,\bm{R}_3)}{2} + \frac{\hbar \bm{\omega}_2}{4}\right) &\nonumber\\
        &\times E\left(\bm{R}_3, \bm{p}_\alpha^{\rm{f}}(\bm{S}_3,\bm{R}_3)\right) F\left(\frac{\bm{S}_2+\bm{S}_3}{2} - \frac{\hbar \bm{s}_2}{4}, \frac{\bm{p}_\alpha^{\rm{f}}(\bm{S}_2,\bm{R}_2) + \bm{p}_\alpha^{\rm{f}}(\bm{S}_3,\bm{R}_3)}{2} - \frac{\hbar \bm{\omega}_2}{4}\right) &\nonumber\\
	&\times  G\left(\bm{\mathcal{X}}  - \frac{\hbar \bm{r}_2}{4}, \bm{\Pi}_1 - \frac{\hbar \bm{\omega}_1}{4}\right) W(2\bm{\mathcal{X}} - \bm{R}_2, 2\bm{\Pi}_1 - \bm{p}_\alpha^{\rm{i}}(\bm{S}_2,\bm{R}_2)).&
\end{flalign}

To introduce the symplectic notation, we add integration over Dirac delta's for the two initial and the two final momenta, resulting in
\begin{flalign}\label{app_G_pair3_sympl}
    O_{CDEFG}^I(t) =\ & \frac{1}{2^{2L}(2\pi)^{4L}}\int \dd X \dd R_{2,3 } \dd S_{2,3} \dd \Delta_{1,2} \ \sum_\alpha \abs{\det\frac{\partial^2 R_\alpha(\bm{S}_2,\bm{R}_2)}{\partial \bm{S}_2 \partial \bm{R}_2}} & \nonumber\\ 
    &\times \abs{\det\frac{\partial^2 R_\alpha(\bm{S}_3,\bm{R}_3)}{\partial \bm{S}_3 \partial \bm{R}_3}} \rme^{\rmi \Delta_1\wedge(R_2-R)} \rme^{\rmi \Delta_2\wedge (S_3 - S_2)/2} &\nonumber \\
    &\times C\left(\frac{X+R_2}{2}+\frac{\hbar \Delta_1}{4}\right)   D\left(\frac{S_2+S_3}{2} +\frac{\hbar \Delta_1}{4}\right) E\left(R_3\right) & \nonumber\\
    &\times  F\left(\frac{S_2+S_3}{2}-\frac{\hbar \Delta_1}{4}\right) G\left(\frac{X+R_2}{2}-\frac{\hbar \Delta_1}{4}\right) &\nonumber\\
    &\times \delta(\bm{\Pi}_2 - \bm{p}^{\rm{i}}_\alpha(\bm{S}_2,\bm{R}_2)) \delta(\bm{\Pi}_3 - \bm{p}^{\rm{i}}_\alpha(\bm{S}_3,\bm{R}_3)) \nonumber\\
    & \times \delta(\bm{P}_2 - \bm{p}^{\rm{f}}_\alpha(\bm{S}_2,\bm{R}_2)) \delta(\bm{P}_3 - \bm{p}^{\rm{f}}_\alpha(\bm{S}_3,\bm{R}_3)) W(2X-R_2) &
\end{flalign}
where part of the expression cannot be put in symplectic formalism. The symplectic vectors are $X = (\bm{\mathcal{X}}, \bm{\Pi}_1)$, $R_2 = (\bm{R}_2, \bm{\Pi}_2)$, $R_3 = (\bm{R}_3, \bm{\Pi}_3)$, $S_2 = (\bm{S}_2, \bm{P}_2)$, $S_3 = (\bm{S}_3, \bm{P}_3)$, $\Delta_1 = (\bm{r}_2, \bm{\omega}_1)$, $\Delta_2 = (\bm{s}_2, \bm{\omega}_2)$, and the four $\delta$'s at the end are introduced to write the expression as phase-space integrals.

\section{Weyl symbol of a squared operator}\label{app_sqopp}
In the following, we will need the expression of the symbol of a squared operator. We will here detail it, as well as its expansion in $\hbar$. 
Applying definition \eqref{app_def_weylSymbol} for a squared operator reads
\begin{flalign}
    (\hat{A}^2)_W(\bm{q},\bm{p}) &= \int \dd\bm{\xi} \bra{\bm{q}+\frac{\bm{\xi}}{2}} \hat{A}^2 \ket{\bm{q}-\frac{\bm{\xi}}{2}} \rme^{-\rmi \bm{p}\cdot \bm{\xi} /\hbar} &\nonumber\\
    &= \int \dd \bm{\xi} \dd \bm{\chi} \matrixel{\bm{q}+\frac{\bm{\xi}}{2}}{\hat{A}}{\bm{\chi}} \matrixel{\bm{\chi}}{\hat{A}}{\bm{q}-\frac{\bm{\xi}}{2}} \rme^{-\rmi \bm{p} \cdot \bm{\xi} / \hbar} &\nonumber\\
    &= \frac{1}{(2\pi \hbar)^{2L}} \int \dd\bm{\xi} \dd\bm{\chi} \dd\bm{p_1} \dd\bm{p_2}\ A\left(\frac{\bm{q}+\bm{\chi}}{2}+\frac{\bm{\xi}}{4}, \bm{p_1}\right) A\left(\frac{\bm{q}+\bm{\chi}}{2}-\frac{\bm{\xi}}{4}, \bm{p_2}\right)& \nonumber \\
    &\quad\times\exp\left[\frac{\rmi}{\hbar} \left( \bm{p_1} \cdot\left(\bm{q}+\frac{\bm{\xi}}{2}-\bm{\chi}\right) + \bm{p_2}\cdot\left(\bm{\chi}-\bm{q} + \frac{\bm{\xi}}{2}\right) - \bm{p} \cdot \bm{\xi} \right)\right].&
\end{flalign}
We then perform a change of variables to go to centre-of-mass and relative coordinates, and we obtain
\begin{flalign} 
    (\hat{A}^2)_W(\bm{q},\bm{p}) =\ &\frac{1}{(2\pi)^{2L}} \int \dd \bm{y} \dd\bm{z} \dd\bm{\Pi} \dd\bm{\omega}\ A\left( \bm{y}+ \frac{\hbar \bm{z}}{4}, \bm{\Pi} + \frac{\hbar \bm{\omega}}{4} \right) A\left( \bm{y}- \frac{\hbar \bm{z}}{4}, \bm{\Pi} - \frac{\hbar \bm{\omega}}{4} \right)&\nonumber\\
 &\times \exp\left[\rmi \left( (\bm{q}-\bm{y})\cdot\bm{\omega} + \bm{z}\cdot(\bm{\Pi} - \bm{p}) \right) \right] &
\end{flalign}
This expression can easily be rewritten in symplectic formalism, using $Y = (\bm{y}, \bm{\Pi})$, $X = (\bm{q},\bm{p})$ and $\Delta = (\bm{z},\bm{\omega})$:
\begin{flalign}
	(\hat{A}^2)_W(X) &= \frac{1}{(2\pi)^{2L} } \int \dd Y \dd\Delta\ \rme^{\rmi \Delta \wedge (Y-X)} A \left(Y+\frac{\hbar \Delta}{4} \right) A \left(Y-\frac{\hbar \Delta}{4} \right) &
\end{flalign}
Then, we make the same assumption we did in Section \ref{sec_semiclassicallimit}: either we consider the operators to be at most quadratic. Doing so leads us to 
\begin{flalign}
    (\hat{A}^2)_W(X) &= \int \frac{\dd Y \dd\Delta}{(2\pi)^{2L} }\ \rme^{\rmi \Delta \wedge (Y-X)} &\nonumber\\
    &\quad \times \left[ A^2 \left(Y\right) + \frac{\hbar^2}{16 } \sum_{k,l=1}^{2L} \Delta_k \Delta_l \left( A(Y) \partial^2_{kl} A(Y) - \partial_k A(Y) \partial_l A(Y)\right)   \right] + \mathcal{O}(\hbar^3) &\nonumber\\
	&=A^2(X) +\frac{\hbar^2}{16}\int \frac{\dd Y \dd \Delta}{(2\pi)^{2L} }\ \rme^{\rmi \Delta \wedge (Y-X)} & \nonumber\\
 & \quad \times \sum_{k,l=1}^{2L}\left[ \Delta_k \Delta_l \left( A(Y) \partial^2_{kl} A(Y) - \partial_k A(Y) \partial_l A (Y)\right)\right]  + \mathcal{O}(\hbar^3) &
\end{flalign}
using $ \int \dd Y \dd \Delta \rme^{\rmi \Delta \wedge (Y-X)}= \left(2 \pi \right)^{2L} \delta \left(Y-X \right)$, where $\partial_k A(Y) = \partial A(Y)/\partial Y_k$ and $\partial^2_{kl} A(Y) = \partial^2 A(Y)/\partial Y_k \partial Y_l$. The next step consists in turning the $\Delta_k$'s of the second term into derivative operators acting on the exponential: 
\begin{eqnarray} \label{eq_deltainDeriv}
    \Delta_k \exp\left[\rmi \Delta \wedge (Y-X)\right)] =  -\frac{\sigma_{\bar{k}}}{\rmi} \frac{\partial}{X_{\bar{k}}} \exp\left[\rmi \Delta \wedge (Y-X)\right].
\end{eqnarray}
As neither the rest of the integrand or the integration bounds depend on $X$, the derivation operators can be taken outside of the integral entirely, which eventually gives
\begin{flalign} 
	(\hat{A}^2)_W(X) &= A^2(X) - \frac{\hbar^2}{8} \sum_{k,l=1}^{2L} \sigma_k \sigma_l  \partial^2_{kl} A(X) \partial^2_{\bar{k} \bar{l}} A(X).
\end{flalign}
Here we use the property:
\begin{flalign}
	\sum_{k=1}^{2L} \sigma_k \frac{\partial f}{\partial Y_k Y_{\bar{k}}} (Y) &= \sum_{k=1}^L \left[] \frac{\partial f}{\partial y_k \partial \Pi_k}(\bm{y},\bm{\Pi}) -  \frac{\partial f}{\partial \Pi_k \partial y_k}(\bm{y},\bm{\Pi})\right] = 0,
\end{flalign}
for an arbitrary twice-differentiable function $f$ of a symplectic vector $Y = (\bm{y}, \bm{\Pi})$, where we have switched from the symplectic to the regular formalism at the first equality. This can also straightforwardly be shown in the symplectic formalism using \mbox{$ \sum_{k=1}^{2L} \sigma_k \partial_k A(Y) = - \sum_{k=1}^{2L} \sigma_{\bar{k}} \partial_{\bar{k}} A(Y) $} since $\sigma_k = -\sigma_{\bar{k}}$.\\
\indent In this development, we assumed that the observable $A$ vanishes at the integration boundaries so that all the expressions that appear are integrable. While this is technically not the case for the standard population operator $A(Y) = (Y_k^2 + Y_{\bar{k}}^2)/2$, we consider this operator to be amended, \textit{e.g.} $A(Y) = \exp(-\varepsilon \sum_{k=1}^{2L} Y_k^2)(Y_k^2 + Y_{\bar{k}}^2)/2$ for sufficiently small $\varepsilon$. This reflects the experimental limitations for the detection of such an observable.

\section{Quasiclassical contributions to the OTOC} \label{sec_app_OTOC_allpairings}
As explained shown by \eqref{eq_app_OTOCFromGeneric}, we can use the expressions of the generic term to build the OTOC, and this for each pairing. This section goes as follows: the contribution of each pairing to the OTOC will be written and the $0^\text{th}$, $1^\text{st}$ and $2^\text{nd}$ order in $\hbar$ will be developped.

\subsection{$V$ term} \label{sec_app_OTOC_Vpairing}
For the $V$ pairing, we get:
\begin{flalign} \label{app_OTOC_pair1}
    C_V(t) &= O_{BA \mathds{1}AB}^V(t) - O_{BABA\mathds{1}}^V(t) - O_{\mathds{1}ABAB}^V(t) + O_{\mathds{1}AB^2 A\mathds{1}}^V(t) &\nonumber\\
    &= \frac{1}{(2\pi)^{4L}} \int \dd X_{1,2,3} d\Delta_{1,2}\ \rme^{\rmi\Delta_1 \wedge (X_2 - X_3)} \rme^{\rmi \Delta_2 \wedge (X_1 - 2X_2 + X_3) }& \nonumber\\
    &\quad \times   A_t\left(X_1 + \frac{\hbar \Delta_2}{4}\right)\ A_t\left(X_1 - \frac{\hbar \Delta_2}{4}\right)    \biggl[\biggr. B\left(X_2 +  \frac{\hbar\Delta_1}{4}\right)\ B\left(X_2 -  \frac{\hbar\Delta_1}{4}\right) &\nonumber\\
     & \quad- B\left(2 X_1 -2 X_2 + X_3\right) \left(B\left(X_2 +  \frac{\hbar\Delta_1}{4}\right)+ B\left(X_2 -  \frac{\hbar\Delta_1}{4}\right)\right) &\nonumber\\
	  & \quad + (\hat{B}^2)_W\left(2 X_1 -2 X_2 + X_3\right) \biggl.\biggr] W(X_3)&
\end{flalign}
where $A_t(X) \equiv A(X_t(X))$ is the symbol of $\hat{A}$ evaluated at the time-evolved coordinate $X_t(X)$. It will abusively be referred to as the time-evolved symbol. The 0$^\text{th}$ order in $\hbar$ corresponds to neglecting any $\hbar$ dependencies. \eqref{app_OTOC_pair1} becomes
\begin{eqnarray} 
	C^V(t) &=&\   \frac{1}{(2\pi)^{4L}} \int \dd X_{1,2,3} \dd\Delta_{1,2}\ \rme^{\rmi\Delta_1 \wedge (X_2 - X_3)} \rme^{\rmi \Delta_2 \wedge (X_1 - 2X_2 + X_3) } \nonumber\\
	&& \times  A_t^2\left(X_1\right)  \biggl[\biggr. B\left(X_2 \right) - B\left(2 X_1 -2 X_2 + X_3\right) \biggl.\biggr] ^2 W(X_3) + \mathcal{O}(\hbar).
\end{eqnarray}
We then perform the integral with respect to $\Delta_1$ and $\Delta_2$, which yields Dirac deltas:
\begin{eqnarray}
	\fl \frac{1}{(2\pi)^{4L}} \int \dd \Delta_{1} \dd \Delta_2\ \rme^{\rmi\Delta_1 \wedge (X_2 - X_3)} \rme^{\rmi \Delta_2 \wedge (X_1 - 2X_2 + X_3) } &=& \delta(X_2-X_3)\delta(X_1 - X_2).
\end{eqnarray}
This will make the square bracket vanish and thus so will the OTOC at the 0$^{\text{th}}$ order.
%
%
The $1^{\text{st}}$ order term in $\hbar$ vanishes as well owing to the $\hbar \longleftrightarrow -\hbar$ symmetry of the expression \eqref{app_OTOC_pair1}.
To evaluate the $2^{\text{nd}}$ order in $\hbar$, one can consider only the expansion up to $\hbar^2$ in the symbol of $\hat{A}$, and in the analogous expansion of the symbol of $\hat{B}$. The two contributions to the OTOC will be labelled $C^V_A$ and $C^V_B$ respectively.\\
\indent For the first term, one can write
\begin{eqnarray} \label{eq_C_V^A}
    C^V_A(t) &=&\frac{\hbar^2}{(2\pi)^{2L}} \int \dd X_{1,2,3} \dd \Delta_2\ e^{i \Delta_2 \wedge (X_1 - X_2) } \sum_{k,l}^{2L} \Delta_{2,k} \Delta_{2,l} \nonumber\\
    &&\times \left[ A_t(X_1) \partial^2_{kl} A_t(X_1) - \partial_k A_t(X_1) \partial_l A_t(X_1)\right]  \nonumber\\
    &&\times \left[ B(X_2)- B\left(2 X_1 - X_2 \right) \right]^2 W(X_2)
\end{eqnarray}
where we have already performed the integration with respect to $\Delta_1$. The procedure is then similar to the one of the symbol of a squared operator, see \eqref{eq_deltainDeriv}. The $\Delta_2$'s will be transformed into derivative operators with respect to components of $X_1$ acting in the exponential, followed this time by an integration by parts that will move said operators to the rest of the integrand. For some generic function $f$ of symplectic vectors $X$ and $Y$ such that $f$ vanishes at the integration boundaries, this translates to
\begin{flalign}
     \int \dd X \dd Y \dd \Delta\ e^{i \Delta \wedge (X - Y) } \Delta_{k}\ f(X,Y) &= \int \dd X \dd Y \dd \Delta\ \frac{\sigma_k}{i} \frac{\partial e^{i \Delta \wedge (X - Y) }}{\partial X_{\bar{k}}}   f(X,Y) &\nonumber\\
     &= \int \dd X \dd Y \dd \Delta\ \frac{-\sigma_k}{i} \delta(X-Y)  \frac{\partial f}{\partial X_{\bar{k}}}(X,Y) &
\end{flalign}
where we integrated by parts to obtain the last line. Applying this to \eqref{eq_C_V^A} leads to 
\begin{eqnarray}
    C^V_A(t)(t) &=& -\frac{\hbar^2}{2} \int \dd X \sum_{k,l=1}^{2L} \sigma_k \sigma_l \partial_{\bar{k}} B(X) \partial_{\bar{l}} B(X) \nonumber\\
 &&\left[ A_t(X) \partial^2_{kl} A_t(X)- \partial_k A_t(X) \partial_l A_t(X)\right]  W(X).
\end{eqnarray}
where the already integrated term from the integration by parts vanishes thanks to the Wigner distribution. A similar approach for $C^V_B$ yields
\begin{eqnarray}
	 C^V_B(t)(t) &=&\frac{\hbar^2}{2} \int \dd X  \sum_{k,l=1}^{2L} \frac{\sigma_k \sigma_l}{\rmi^2} \partial_{\bar{k}} B(X) \partial_{\bar{l}} B(X)  \nonumber\\
 && \times \left[- A_t (X) \partial^2_{kl} A_t(X) - \partial_k A_t(X) \partial_l A_t (X)\right]  W(X).
\end{eqnarray}
The classical OTOC of the $V$ pairing $C^V_{cl}$, \textit{i.e.} the first non-vanishing contribution to the OTOC, is written
\begin{eqnarray}
	C^V_{cl}(t) &= C^V_A(t) + C^V_B(t) &\nonumber\\
	&= \hbar^2 \int \dd X \sum_{k,l=1}^{2L} \sigma_k \sigma_l\ \partial_k A_t(X) \partial_l A_t(X)   \partial_{\bar{k}} B(X) \partial_{\bar{l}} B(X)  W(X) &\nonumber \\
	&= \hbar^2 \int \dd X \left\{ A_t(X), B(X) \right\}^2 W(X).
\end{eqnarray}

\subsection{$\Lambda$ term} \label{sec_app_OTOC_Lambdapairing}
Once again, we can then construct the OTOC on the basis of the generic term:
\begin{flalign} \label{app_OTOC_pair2}
	C^\Lambda(t)(t) &= O_{BA\mathds{1}AB}^\Lambda(t) - O_{BABA\mathds{1}}^\Lambda(t) - O_{\mathds{1}ABAB}^\Lambda(t) + O_{\mathds{1}AB^2 A\mathds{1}}^\Lambda(t) &\nonumber\\
    &=\frac{1}{(2\pi \hbar)^{4L}} \int \dd X \dd S_{23} \dd \Delta_{12}\ \rme^{\rmi \Delta_1\wedge(X_{-t}(2S_2-S_3)-X)} \rme^{\rmi \Delta_2\wedge (S_3 - S_2)} &\nonumber\\
    &\quad \times A\left(S_2+ \frac{\hbar \Delta_2}{4}\right) A\left(S_2- \frac{\hbar \Delta_2}{4}\right) \bigg[\biggr. B\left(X+\frac{\hbar \Delta_1}{4}\right) B\left(X-\frac{\hbar \Delta_1}{4}\right) &\nonumber\\
	& \quad -B\left(X_{-t}(S_3)\right)\left[B\left(X+\frac{\hbar \Delta_1}{4}\right) + B\left(X-\frac{\hbar \Delta_1}{4}\right)\right] & \nonumber\\
	&\quad +\left(\hat{B}^2\right)_W\left(X_{-t}(S_3)\right)\biggl.\biggr]  W\left(2X-X_{-t}(2S_2-S_3)\right).&
\end{flalign}
We'll follow the same procedure as before, \textit{i.e.} we'll look at the $0^\text{th}$, $1^\text{st}$ and $2^\text{nd}$ order in $\hbar$. The $0^{\text{th}}$ and $1^{\text{st}}$ order vanish for the same reason as in the $V$ pairing. The first non-vanishing contribution to the out-of-time-ordered correlator is again the $2^{\text{nd}}$. As the lower orders are shown to be zero, one can once again look separately at the expansion of the $A$ and $B$ symbols.\\
\indent The second-order expansion of the $A$ symbols yield
\begin{eqnarray} 
    C^\Lambda_A(t) &=\frac{\hbar^2}{16(2\pi)^{2L}}  \int \dd X \dd S_{23} \dd \Delta_{23}\ \rme^{\rmi \Delta_1\wedge(X_{-t}(2S_2-S_3)-X)} \rme^{\rmi \Delta_2\wedge (S_3 - S_2)} &\nonumber\\
    &\quad  \sum_{i,j=1}^{2L} \Delta_{2i} \Delta_{2j} \left[ A(S_2) \partial^2_{ij} A(S_2) - \partial_i A(S_2) \partial_j A(S_2)\right] &\nonumber\\
    & \quad\times\left[ B\left(X\right) -B\left(X_{-t}(S_3)\right)\right]^2  W\left(2X-X_{-t}(2S_2-S_3)\right).&
\end{eqnarray}
One can perform the integration with respect to $\Delta_1$ and transform the perturbation variables $\Delta_2$'s into derivation operators acting on the exponential, remove them from said exponential by integrating by part, and end up with
\begin{eqnarray} 
    C^\Lambda_A(t) &= \frac{\hbar^2}{4} \int \dd S \sum_{i,j=1}^{2L} \frac{\sigma_i \sigma_j}{\rmi^2} \left[ A(S) \partial^2_{ij} A(S) - \partial_i A(S)\partial_j A(S)\right] &\nonumber\\
    &\quad \times \frac{\partial B}{\partial S_{\bar{i}}}(X_{-t}(S))\frac{\partial B}{\partial S_{\bar{j}}}(X_{-t}(S))  W\left(X_{-t}(S)\right). &
\end{eqnarray}
For the perturbation in the $B$ symbols, one obtains at first
\begin{flalign}
    C^\Lambda_B(t) =\ &\frac{\hbar^2}{16 {(2\pi)^{2L}}} \int \dd X \dd S_{23} \dd \Delta_{23}\ \rme^{\rmi \Delta_1\wedge(X_{-t}(2S_2-S_3)-X)} \rme^{\rmi \Delta_2\wedge (S_3 - S_2)} A^2(S_2)  &\nonumber \\
    & \quad \times \sum_{i,j=1}^{2L} \Delta_{1i} \Delta_{1j} \left[ B(X) \partial^2_{ij} B(X) - \partial_i B(X) \partial_j B  +B(X_{-t}(S_3)) \partial^2_{ij} B(X) \right]  &\nonumber\\
    & \quad \times W\left(2X-X_{-t}(2S_2-S_3)\right) &\nonumber\\
    & + \frac{\hbar^2}{8} \int \dd X\ A_t^2(X)\sum_{i,j=1}^{2L} \frac{\sigma_i \sigma_j}{\rmi^2}  \partial^2_{\bar{i} \bar{j}} B(X) \partial^2_{ij} B(X) W(X)&
\end{flalign}
and when following the same guidelines as above, ends up with
\begin{flalign} 
	C^\Lambda_B(t) &= \frac{\hbar^2}{4}  \int \dd X \sum_{i,j=1}^{2L} \frac{\sigma_i \sigma_j}{\rmi^2}  A^2_t(X) \biggl[\biggr. \partial^2_{\bar{i} \bar{j}} B(X) \partial^2_{ij} B(X)  W(X) &\nonumber \\
    &\quad  + 2  \partial_{\bar{i}} B(X) \partial^2_{ij} B(X)  \partial W_{\bar{j}}(X) + \partial_{\bar{i}} B(X) \partial_j B(X) \partial^2_{i \bar{j}} W(X)  \biggl.\biggr] &\\
    &= \frac{\hbar^2}{4}  \int \dd X \sum_{i,j=1}^{2L} \sigma_i \sigma_j  A_t^2(X) \frac{\partial^2}{\partial X_{\bar{i}} \partial X_{\bar{j}}} \left[ \partial_i B(X) \partial_j B(X) W(X) \right].&
\end{flalign}

\indent When integrating by parts this last term and putting everything together, we end up with the classical OTOC of the $\Lambda$ pairing
\begin{flalign}
    C^\Lambda_{cl}(t) &= C^\Lambda_A(t) + C^\Lambda_B(t) &\nonumber\\
    &=\int \dd S\ \hbar^2 \sum_{i,j} \sigma_i \sigma_j \partial_i A(S) \partial_j A(S) \frac{\partial B}{\partial S_{\bar{i}}}(X_{-t}(S)) \frac{\partial B}{\partial S_{\bar{j}}}(X_{-t}(S))  W(X_{-t}(S)) &\nonumber\\
    &\quad - \frac{\hbar^2}{2} \int \dd S \sum_{i,j=1}^{2L} \sigma_i \sigma_j  A(S) \partial^2_{ij} A \frac{\partial B}{\partial S_{\bar{i}}}(X_{-t}(S)) \frac{\partial B}{\partial S_{\bar{j}}}(X_{-t}(S))  W(X_{-t}(S)) &\nonumber\\
    &\quad + \frac{\hbar^2}{2} \int \dd X \sum_{i,j=1}^{2L} \sigma_i \sigma_j  A_t \frac{\partial^2 A_t}{\partial X_{\bar{i}} \partial X_{\bar{j}}}(X)   \partial_i B(X) \partial_j B(X)  W(X) &\nonumber\\
	&=\hbar^2 \int \dd X\ \left\{ A(S(X)), B(X)\right\}^2 W(X) \nonumber\\
    &\quad - \frac{\hbar^2}{2} \int \dd S \sum_{i,j=1}^{2L} \sigma_i \sigma_j  A(S) \partial^2_{ij} A(S) \frac{\partial B}{\partial S_{\bar{i}}} (X_{-t}(S)) \frac{\partial B}{\partial S_{\bar{j}}}(X_{-t}(S))  W(X_{-t}(S)) &\nonumber\\
    &\quad + \frac{\hbar^2}{2} \int \dd X \sum_{i,j=1}^{2L} \sigma_i \sigma_j  A(S) \frac{\partial^2 A}{\partial X_{\bar{i}} \partial X_{\bar{j}}} (X) \partial_i B(X) \partial_j B(X) W(X)&
\end{flalign}
using the fact that the Poisson bracket is  canonically invariant. In other words, the contribution of the second pairing to the short time OTOC is again a Poisson bracket plus some other contributions, but fortunately we will show that the latter vanishes.

\subsection{$I$ term} \label{sec_app_OTOC_Ipairing}
\begin{figure}[H]
		\centerline{
			\includegraphics[width=\textwidth, angle=0, trim={0cm 0cm 0cm 0cm}, clip]{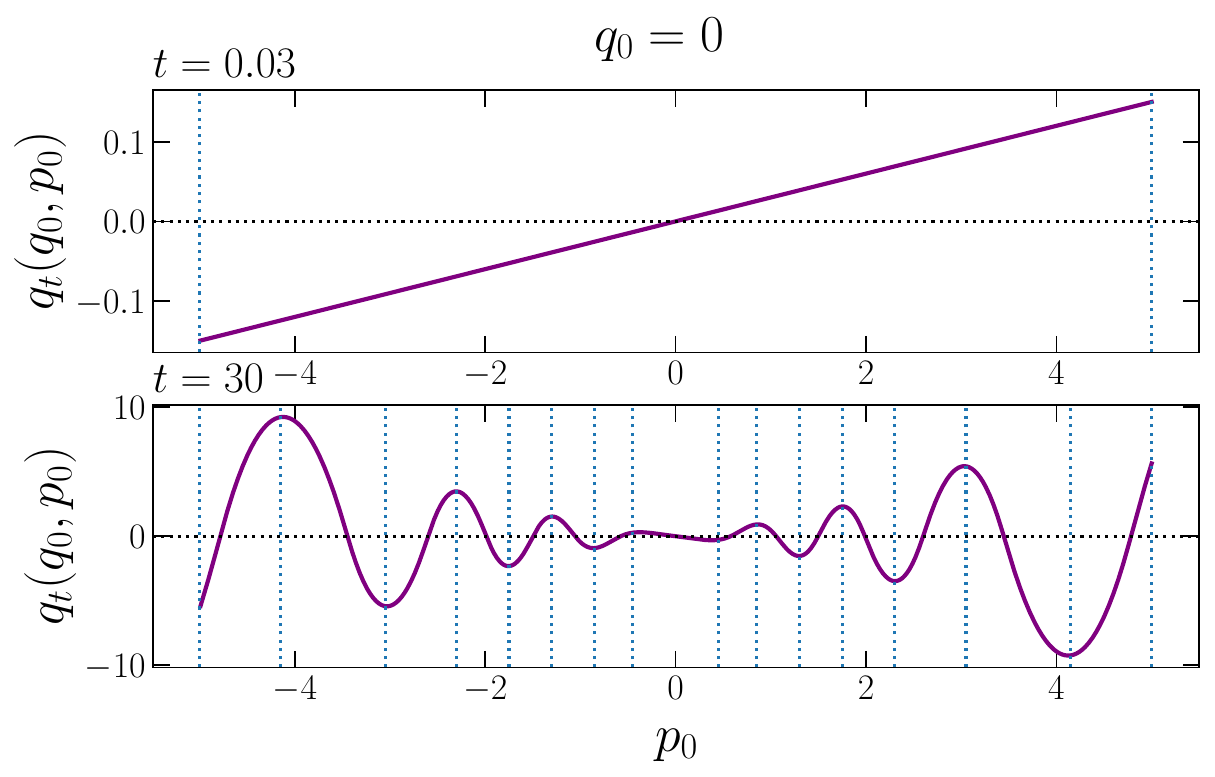}
		}
		\captionsetup{justification=justified}
		\caption{Final position as a function of initial momentum, at a fixed initial position $q_0=0$, for the classical one-degree-of-freedom system governed by the Hamiltonian $H(q,p) =  p^2/2 + \sqrt{1+q^2}$. The vertical dotted lines delimit the different families of trajectories departing at $q_0=0$ and ending near $q_t=0$ again, after the evolution time $t=0.03$ (resp. $t=30$) for the top (resp. bottom) panel. In the top panel, representing the short-time situation, there is only one possible family of trajectories, \textit{i.e.} meaning that inside \eqref{app_OTOC_pair3_2}, $\delta_{\alpha_2, \alpha_3}=1$. The bottom panel shows a situation at later time, where there are multiple families fulfilling the aforementioned conditions. The consequence is that in \eqref{app_OTOC_pair3_2}, the phase-space integrals cannot be performed independently. To obtain a non-vanishing contribution to the integral, the trajectories must belong to the same family during the whole evolution, \textit{i.e.} the initial condition must be in the same region delimited by the vertical dotted lines. As time increases, this region, and thus the integration domain, becomes smaller, leading to a smaller contribution to the OTOC.}
		\label{app_fig_qt}
\end{figure}

When considering both the $V$ and $\Lambda$ pairing, we are actually overcounting one situation: when all four trajectories are near one another, \textit{i.e.} $\alpha \sim \beta \sim \gamma \sim \delta$. Therefore, we have to subtract once this contribution from the other two.
Using the same set of variables as for the $\Lambda$ pairing seems far more suitable for the derivation. One can thus follow the same step as for the last pairing and obtain
\begin{flalign} \label{app_OTOC_pair3_1}
    C^I(t) &= O_{BA\mathds{1}AB}^I(t) - O_{BABA\mathds{1}}^I(t) - O_{\mathds{1}ABAB}^I(t) + O_{\mathds{1}AB^2 A\mathds{1}}^I(t) & \nonumber\\
        &=\frac{1}{(2\pi )^{4L} 2^{2L}} \int \dd X \dd R_{2,3} \dd S_{2,3} \dd \Delta_{12} \sum_\alpha \abs{\det\frac{\partial^2 R_\alpha(\bm{S}_2,\bm{R}_2)}{\partial \bm{S}_2 \partial \bm{R}_2}}  &\nonumber\\
	&\quad \times \abs{\det\frac{\partial^2 R_\alpha(\bm{S}_3,\bm{R}_3)}{\partial \bm{S}_3 \partial \bm{R}_3}}  \rme^{\rmi \Delta_1 \wedge (R_2-X)} \rme^{\rmi \Delta_2 \wedge (S_3-S_2)/2}  A\left(\frac{S_2+S_3}{2} + \frac{\hbar \Delta_2}{4}\right)&\nonumber \\
	&\quad \times   A\left(\frac{S_2+S_3}{2}- \frac{\hbar \Delta_2}{4}\right)  \biggl[\biggr. B\left(\frac{X+R_2}{2}+\frac{\hbar \Delta_1}{4}\right) B\left(\frac{X+R_2}{2}-\frac{\hbar \Delta_1}{4}\right) &\nonumber\\ 
	&\quad  -B\left(R_3\right)\left(B\left(\frac{X+R_2}{2}+\frac{\hbar \Delta_1}{4}\right) + B\left(\frac{X+R_2}{2}-\frac{\hbar \Delta_1}{4}\right)\right) +\left(\hat{B}^2\right)_W\left(R_3\right) \biggl.\biggr] &\nonumber\\
    &\quad \times \delta(\bm{\Pi}_2 - \bm{p}^{\rm{i}}_\alpha(\bm{S}_2,\bm{R}_2)) \delta(\bm{\Pi}_3 - \bm{p}^{\rm{i}}_\alpha(\bm{S}_3,\bm{R}_3))  \delta(\bm{P}_2 - \bm{p}^{\rm{f}}_\alpha(\bm{S}_2,\bm{R}_2))  \nonumber\\
    & \quad \times \delta(\bm{P}_3 - \bm{p}^{\rm{f}}_\alpha(\bm{S}_3,\bm{R}_3)) W\left(X\right).&
\end{flalign}
There is however one caveat. Whereas in the previous pairings, a double sum on the classically allowed trajectories remained, here there is only one as only one family of paths is considered. Let us first transform the previous expression. The Dirac $\delta$ in combination with a function $f$ in multiple dimensions is given by
\begin{flalign}\label{app_composite_delta}
	\delta(\bm{y}-f(\bm{x})) = \sum_i \frac{\delta(\bm{x}-\bm{x}_i)}{\abs{\det \frac{\dd f}{\dd \bm{x}}(\bm{x}_i) }}
\end{flalign}
where the sum is on all $\bm{x}_i$ that satisfy $\bm{y}-f(\bm{x})=0$. To transpose it to our case that involves several variables, we simply have to choose one of the variable as being the $\bm{x}$ in the above expression, and the other remains constant. This can be done by choosing which integral to perform first, as the Dirac $\delta$ should be understood in this context. This yields:
\begin{flalign} \label{app_delta_composition}
     \prod_{j=2}^3 \abs{\det\frac{\partial^2 R_\alpha(\bm{S}_j,\bm{R}_j)}{\partial \bm{S}_j \partial \bm{R}_j}} &  \delta\left(\bm{\Pi}_j - \bm{p}_\alpha^\rmi(\bm{S}_j,\bm{R}_j)\right) \delta(\bm{P}_j - \bm{p}_\alpha^{\rm{f}}(\bm{S}_j,\bm{R}_j))&\nonumber\\
     =\ \prod_{j=2}^3  &\delta(\bm{S}_j-\bm{R}_t(\bm{R}_j, \bm{\Pi}_j))  \delta(\bm{P}_j - \bm{P}_t(\bm{R}_j,\bm{\Pi}_j)) \delta_{\alpha,\alpha_j}& \nonumber\\
     =\ \prod_{j=2}^3  &\delta(S_j - X_t(R_j)) \delta_{\alpha \alpha_j} &
\end{flalign}
where we switched to the symplectic formalism for the last equality. In this case the sum in \eqref{app_composite_delta} contains only one element since there is only one zero to the argument of the Dirac delta. The index $\alpha_j$ labels the family of trajectories $\bm{R}_j \to \bm{S}_j$ for which $\bm{\Pi}_j$ lies in the image set of $\bm{p}_{\alpha_j}^{\rm{i}}$. Summing over $\alpha$ on the individual factors of expression \eqref{app_delta_composition} yields $\delta(S_j-X_t(R_j))$. Here, however, the sum is performed over the whole product, according to equation \eqref{app_OTOC_pair3_1}, thus yielding $\delta_{\alpha_2, \alpha_3} \prod_{j=2}^3 \delta(S_j - X_t(R_j))$. This means that the two involved trajectory families, labeled by $\alpha_2$ and $\alpha_3$, must be identical in order for this sum to yield a nonzero value. One therefore obtains
\begin{flalign}
    C^I(t) &=\frac{1}{(2\pi )^{4L} 2^{2L}} \int \dd X \dd R_{2,3} \dd S_{2,3} \dd \Delta_{12} \ \delta_{\alpha_2,\alpha_3} \delta(S_2 - X_t(R_2)) \delta(S_3 - X_t(R_3))  &\nonumber\\
	&\quad \times \rme^{\rmi \Delta_1 \wedge (R_2-X)} \rme^{\rmi \Delta_2 \wedge (S_3-S_2)/2}  A\left(\frac{S_2+S_3}{2} + \frac{\hbar \Delta_2}{4}\right) A\left(\frac{S_2+S_3}{2}- \frac{\hbar \Delta_2}{4}\right) &\nonumber \\
	&\quad \times  \biggl[\biggr. B\left(\frac{X+R_2}{2}+\frac{\hbar \Delta_1}{4}\right) B\left(\frac{X+R_2}{2}-\frac{\hbar \Delta_1}{4}\right) &\nonumber\\ 
	&\quad  -B\left(R_3\right)\left(B\left(\frac{X+R_2}{2}+\frac{\hbar \Delta_1}{4}\right) + B\left(\frac{X+R_2}{2}-\frac{\hbar \Delta_1}{4}\right)\right) +\left(\hat{B}^2\right)_W\left(R_3\right) \biggl.\biggr] & \nonumber\\
    & \quad \times W\left(X\right) & \nonumber\\
    &=\frac{1}{(2\pi )^{4L} 2^{2L}} \int \dd X \dd R_{2,3} \dd \Delta_{12} \ \delta_{\alpha_2,\alpha_3} \rme^{\rmi \Delta_1 \wedge (R_2-X)} \rme^{\rmi \Delta_2 \wedge (X_t(R_3)-X_t(R_2))/2}   &\nonumber\\
	&\quad \times A\left(\frac{X_t(R_2)+X_t(R_3)}{2} + \frac{\hbar \Delta_2}{4}\right) A\left(\frac{X_t(R_2)+X_t(R_3)}{2}- \frac{\hbar \Delta_2}{4}\right) &\nonumber \\
	&\quad \times  \biggl[\biggr. B\left(\frac{X+R_2}{2}+\frac{\hbar \Delta_1}{4}\right) B\left(\frac{X+R_2}{2}-\frac{\hbar \Delta_1}{4}\right) &\nonumber\\ 
	&\quad  -B\left(R_3\right)\left(B\left(\frac{X+R_2}{2}+\frac{\hbar \Delta_1}{4}\right) + B\left(\frac{X+R_2}{2}-\frac{\hbar \Delta_1}{4}\right)\right) +\left(\hat{B}^2\right)_W\left(R_3\right) \biggl.\biggr]& \nonumber\\ 
    & \quad \times W\left(X\right) & \nonumber
    \end{flalign}
    \begin{flalign} \label{app_OTOC_pair3_2}
    &\qquad = \frac{1}{(2\pi )^{4L} } \int \dd X \dd S_{2,3} \dd \Delta_{12} \ \delta_{\alpha_2,\alpha_3} \rme^{\rmi \Delta_1 \wedge (X_{-t}(2S_2-S_3)-X)} \rme^{\rmi \Delta_2 \wedge (S_3-S_2)}   &\nonumber\\
	&\qquad \quad \times A\left(S_2 + \frac{\hbar \Delta_2}{4}\right) A\left(S_2- \frac{\hbar \Delta_2}{4}\right)  \biggl[\biggr. B\left(X+\frac{\hbar \Delta_1}{4}\right) B\left(X-\frac{\hbar \Delta_1}{4}\right) &\nonumber\\ 
	&\qquad \quad  -B\left(X_{-t}(S_3)\right)\left(B\left(X+\frac{\hbar \Delta_1}{4}\right) + B\left(X-\frac{\hbar \Delta_1}{4}\right)\right) +\left(\hat{B}^2\right)_W\left(X_{-t}(S_3)\right) \biggl.\biggr] &\nonumber\\
    &\qquad \quad \times W\left(2X-X_{-t}(2S_2-S_3)\right) &
\end{flalign}
where we did changes of variables to simplify the arguments of the symbols. Note that the expression is now identical to \eqref{app_OTOC_pair2}, except for the additional Kronecker delta. This additional Kronecker will be $1$ for fixed time as $\hbar \to 0$. Nonetheless, as time gets larger, the initial conditions must be closer to each other in order to contribute, meaning that the contribution to \eqref{app_OTOC_pair3_2} will get smaller, down the point where it vanishes. Physically, this should be understood as the fact that all trajectories remaining in the vicinity of each other during the whole time evolution becomes increasingly unlikely. This can be further explained using figure \ref{app_fig_qt}. In the short-time situation, there is only one family of trajectories, \textit{i.e.} only one set of initial conditions that yields the desired final conditions. At later times, the system can start from the same initial position but different momenta, and end up to the desired final conditions. However, these different momenta means that the trajectories belong to a different family, in other words have a different index $\alpha$, and thus don't contribute to the integral. \\
\indent This means that for short time, we can follow exactly the same derivations as for the $\Lambda$ pairing, and since \eqref{app_OTOC_pair3_2} contributes with opposite sign to the OTOC, only the $V$ pairing contributes for short time. 
For later times, as it was stated above, only the first two pairings are expected to contribute for a quasiclassical long-time value.

\section*{References}
\bibliographystyle{iopart-num}
\bibliography{bibliography}

\end{document}